\documentclass[aps,prx,10pt,twocolumn,superscriptaddress,amsmath,amssymb,longbibliography,floatfix]{revtex4-2}

\usepackage{graphicx}
\usepackage{amsmath,amssymb,amsthm}
\usepackage{mathtools}
\usepackage{tikz}
\usepackage{braket}
\usepackage{xcolor}
\usepackage{algpseudocode}
\usepackage{booktabs}
\usepackage{longtable}
\usepackage{tikz-3dplot}
\usepackage{hyperref}
\usepackage{csquotes}

\newcounter{algocounter}
\newcommand{\algstart}[2]{
  \refstepcounter{algocounter}%
  \hrule\vspace{4pt}%
  \noindent\textbf{Algorithm~\thealgocounter:} #1\label{#2}%
  \vspace{2pt}\hrule\vspace{4pt}%
}
\newcommand{\algfinish}{\vspace{2pt}\hrule}
\newcommand{\inlinefigcaption}[2]{%
  \refstepcounter{figure}%
  \label{#1}%
  \par\smallskip
  {\small\noindent\textbf{FIG. \thefigure.} #2\par}%
  \medskip
}
\newtheorem*{theorem*}{Theorem}
\newtheorem*{proposition*}{Proposition}
\newtheorem*{corollary*}{Corollary}
\newtheorem*{definition*}{Definition}

\newcommand{\phys}{\mathrm{phys}}
\newcommand{\logi}{\mathrm{log}}

\newcommand{\ps}[1]{#1^{\mathrm{phys}}}

\newcommand{\Tr}{\mathrm{Tr}}
\newcommand{\CCZ}{\mathrm{CCZ}}
\newcommand{\CCCZ}{\mathrm{CCCZ}}
\newcommand{\CZ}{\mathrm{CZ}}

\newcommand{\CS}{\mathrm{CS}}
\newcommand{\CCS}{\mathrm{CCS}}
\newcommand{\CT}{\mathrm{CT}}

\newcommand{\CNOT}{\mathrm{CNOT}}
\newcommand{\SWAP}{\mathrm{SWAP}}
\newcommand{\Aut}{\mathrm{Aut}}

\newcommand{\Fbb}{\mathbb{F}}

\definecolor{faceX1}{RGB}{100, 149, 237} 
\definecolor{faceX2}{RGB}{255, 99, 71}   
\definecolor{faceX3}{RGB}{60, 179, 113}  
\usepackage{amscd}
\begin{document}

\title{Efficiently simulable quantum circuits with large entanglement, magic, and non-Gaussianity via code-compiled tensor networks}

\author{Aydin Deger}
\thanks{These authors contributed equally to this work. Corresponding author: \url{aydin.deger@physics.ox.ac.uk}.}
\affiliation{Department of Physics, Clarendon Laboratory, University of Oxford, Parks Road, Oxford OX1 3PU, United Kingdom}
\author{Stergios Koutsioumpas}
\thanks{These authors contributed equally to this work. Corresponding author: \url{aydin.deger@physics.ox.ac.uk}.}
\affiliation{School of Informatics, The University of Edinburgh, United Kingdom}
\author{Mark Webster}
\affiliation{Department of Physics \& Astronomy, University College London, London, WC1E 6BT, United Kingdom}
\author{Hasan Sayginel}
\affiliation{Department of Physics \& Astronomy, University College London, London, WC1E 6BT, United Kingdom}
\affiliation{National Physical Laboratory, Teddington, TW11 0LW, United Kingdom}
\author{Joschka Roffe}
\affiliation{School of Informatics, The University of Edinburgh, United Kingdom}
\author{Dan E. Browne}
\affiliation{Department of Physics \& Astronomy, University College London, London, WC1E 6BT, United Kingdom}

\date{\today}

\begin{abstract}
We introduce a family of quantum circuits that possess standard indicators of classical simulation hardness, including high entanglement entropy, magic, and non-Gaussianity, yet admit efficient classical simulation via matrix product states (MPS).
Our construction relies on logical circuits of high-rate Calderbank--Shor--Steane (CSS) codes with enhanced symmetries. Using code automorphisms and transversal diagonal gates from higher levels of the Clifford hierarchy, we realize nonlocal logical Clifford and non-Clifford gates. This framework shows how error-correcting codes can compile complex logical circuits into simple physical operations.
The simulation exploits two structural properties: (i)~diagonal transversal gates do not increase MPS bond dimension, and (ii)~permutations can be tracked classically via on-the-fly relabeling, avoiding costly SWAP networks.
Unlike Clifford or matchgate simulation, our method accepts a broad class of initial states, including dense entangled, magic, and non-Gaussian inputs, provided the encoded state retains an efficient MPS representation.
Besides MPS, we release an exact phase-polynomial simulation backend for monomial subfamilies. Its cost is controlled not by entanglement growth, but by the higher-degree phase terms left beyond the quadratic Gaussian sum.
We demonstrate the method on an infinite polar CSS code family and show that bond dimension remains bounded by the encoding cost, independent of logical circuit depth.
Our results show that for some circuit families, standard resource measures are individually insufficient to indicate simulation hardness. As a near-term application, we use the compiled MPS as a classical reference for direct fidelity estimation of a quantum device running nontrivial logical circuits. Perfect Pauli sampling on the encoded reference, together with a Clifford pushback through the known encoder, provides the ideal expectation values, so that the logical output fidelity can be estimated from local Pauli readout alone without the need for costly state tomography.
\end{abstract}

\maketitle

\begin{figure*}[!t]
    \centering
\includegraphics[width=1.\linewidth]{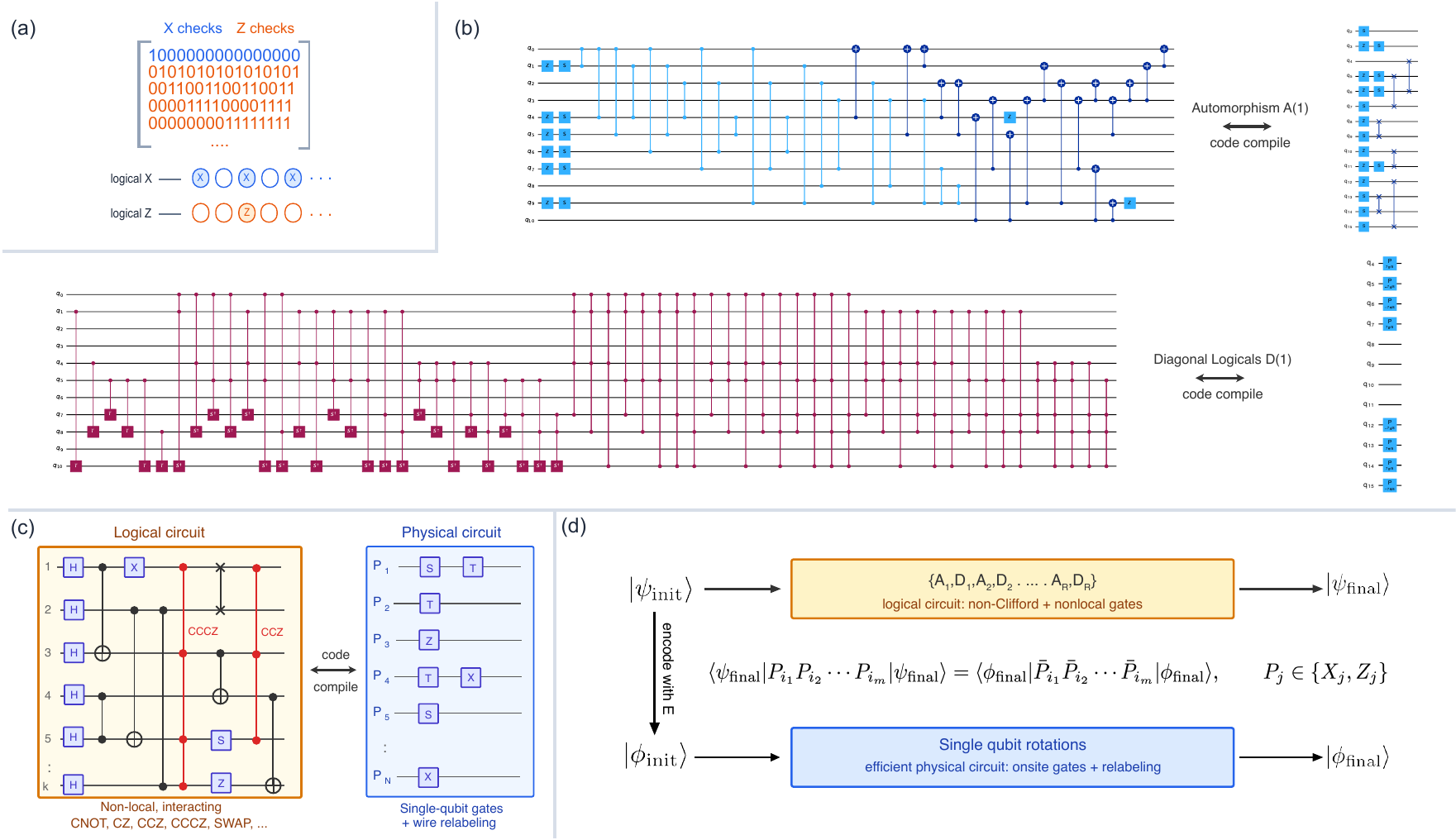}
    \caption{Code-compilation framework.  (a) $X$- and $Z$-check matrices, together with chosen logical $\bar X$ and $\bar Z$.
(b) Two examples of matched physical/logical pairs. Top row: a single physical Clifford automorphism layer compiles to a dense, non-local logical Clifford circuit. Bottom row: a single physical transversal diagonal phase layer compiles to a non-trivial logical diagonal action containing two- and three-body phase terms.
(c) Illustrative example: we randomly compose automorphism and diagonal matched pairs from the precomputed library. The resulting logical circuit (left, on $k$ logical qubits) is highly nonlocal and contains both Clifford and non-Clifford gates such as $\CNOT$, $\CZ$, $\CS$, $\CCZ$, $\CCCZ$, and $\SWAP$, while the corresponding compiled physical circuit (right, on $N$ physical qubits) is just single-qubit operations together with relabelings.
(d) Schematic of the equivalence used by the simulator. The top path shows the logical circuit on \(k\)
qubits. The bottom path first encodes into \(N\) physical qubits and then
applies the physical blocks. The physical circuit uses only onsite gates
and relabeling, but reproduces the same logical observables:
\(\langle O\rangle_{\psi_{\rm final}}
=
\langle \bar O\rangle_{\phi_{\rm final}}\), where \(\bar O\) is the encoded
representative of the logical observable \(O\).
}
    \label{fig:group}
\end{figure*}

\section{Introduction}
When does a quantum circuit become hard to simulate classically? This question lies at the heart of quantum
advantage~\cite{daley_practical_2022, hangleiter2026}. To claim that a quantum device outperforms any classical competitor, one must understand where efficient classical simulation ends and genuine quantum hardness begins.

Three canonical families of circuits are known to admit efficient classical simulation, each through a different mechanism. Clifford circuits on stabilizer-state inputs are handled by stabilizer tableaus via the Gottesman--Knill theorem~\cite{gottesman1998,aaronson2004}. Matchgate circuits on Gaussian states reduce to free-fermion dynamics and admit Pfaffian formulas~\cite{terhal2002,jozsa2008,valiant2002,jozsa_miyake_strelchuk_2015}. Circuits whose entanglement remains bounded throughout the evolution are captured by matrix product states (MPS) at cost polynomial in the bond dimension~\cite{vidal2003,schollwock2011}. More general tensor-network contraction methods can also exploit circuit geometry to obtain subexponential simulation bounds for finite-range circuits~\cite{wahl_strelchuk_2023}. The boundaries of each family are sharp. A $T$ gate added to Cliffords, or non-nearest-neighbour matchgates added to matchgates, already lifts each model to universality, and generic dynamics drive the entanglement to volume law within a depth of order $N$. Methods that push past these boundaries by allowing a controlled budget of universality-enabling resources, such as stabilizer-rank simulation of Clifford+$T$ circuits~\cite{bravyi2016,bravyi2018cliffordt} and non-Gaussian extensions of matchgate circuits~\cite{mocherla2023} and parity-preserving fermionic circuit simulators~\cite{wille_strelchuk_2025}, retain efficiency only when the budget itself grows slowly with system size. A circuit that exhibits volume-law entanglement, sizeable nonstabilizerness, and non-Gaussian correlations together is therefore taken to lie outside the reach of all of these methods.

Here we show that the simulability landscape is richer than these three families suggest.
We introduce \emph{code-compiled quantum circuits} (CCQCs). A CCQC starts from the CSS code~\cite{calderbank1996,steane1996} data shown schematically in Fig.~\ref{fig:group}(a): $X$- and $Z$-checks, together with chosen logical Pauli representatives, define the encoded degrees of freedom. The second ingredient is a list of physical/logical operations. Code automorphisms~\cite{681315,grassl_leveraging_2013,sayginel2024} provide the logical Clifford sector, often inducing dense $\CNOT$, $\SWAP$, Pauli, and $S$ logical operations from simple physical qubit relabelings. Transversal diagonal operators from higher levels of the Clifford hierarchy~\cite{cui2017,Webster_2023,jain_high-distance_2024,eastin2009,koutsioumpas_smallest_2022} provide the non-Clifford sector, including logical $T$, controlled-$S^\dagger$, and $\CCZ$ entries at level $3$, and level-$4$ or higher entries such as $\CT$, $\CCS$, and $\CCCZ$. These pairs are the basic dictionary of the construction. As illustrated in Fig.~\ref{fig:group}(b), a simple physical relabeling and transversal phase layer can act as a nonlocal logical Clifford or a many-body logical non-Clifford diagonal gate. The same operation can therefore have two sharply different descriptions. The resulting catalogs are structured and non-universal, as expected from the restrictions on transversal gates \cite{eastin2009}.

We use this mixed catalog throughout. Composing the pairs gives the contrast shown in Fig.~\ref{fig:group}(c). At the logical level one sees a nonlocal circuit on $k$ encoded qubits, containing both Clifford and non-Clifford gates. At the physical level, after the encoder $E$, the same circuit is only an onsite phase on $N$ physical qubits. The logical dynamics can generate volume-law entanglement, large nonstabilizerness, and departure from the Gaussian manifold, while the compiled physical dynamics retains this simple layer structure.

The classical simulation follows the physical path in Fig.~\ref{fig:group}(d), and its cost is controlled by two properties of the compiled layers. Single-qubit diagonal gates leave the Schmidt rank at every cut of an MPS invariant, and qubit permutations are tracked classically through a qubit-to-site lookup table rather than realized as SWAP networks. Neither operation modifies the stored tensors. Once the encoder has produced the initial encoded MPS, no later logical gate has to be applied as a nonlocal tensor-network operation. Bond-dimension growth is therefore confined to the encoder. This framework also admits a substantially broader input class than Clifford or matchgate simulation, including product states, entangled states, and states with appreciable magic or non-Gaussianity, provided the encoded MPS description remains efficient. The circuits can be made arbitrarily deep without any further bond-dimension growth.

This compiled structure also exposes a second, complementary simulation backend. In the MPS setting, the cost is charged to entanglement growth through the encoded bond dimension.
We also release a complementary exact phase-polynomial backend, \textsc{PhasePoly.jl}~\cite{phasepoly}, where the computational complexity is not governed by entanglement growth but to the algebraic complexity of the phase polynomial. At the logical level, level-$4$ phase terms such as $\CCCZ$ leave cubic Boolean polynomials in the relevant finite difference, and exact evaluation of such cubic gaps is \#P-hard in general~\cite{montanaro2017lowdegree, maslov_fast_2024}. On the other hand, compilation via QEC changes the object being evaluated. The physical circuit may contain a different phase polynomial, on more variables but with lower-degree local phase terms, together with the permutation structure induced by the encoded Clifford/CNOT layers. The two backends therefore cross-check one another but cover different regimes. The deep, dense, magic-rich circuits emphasised below are simulated with the MPS backend, while \textsc{PhasePoly.jl} is most useful when only a small higher-degree remainder is left in the physical circuits.

We study the construction in detail for an infinite family of polar CSS codes~\cite{7208851,PhysRevLett.113.030501,renes2012efficient}, which combines the required structural properties with an efficient encoding. The peak bond dimension reached during encoding is the only quantity that controls the simulation cost, after which all compiled layers preserve the stored bond dimension.
For the polar family with $N$ qubits and stabilizer input states, the peak bond dimension satisfies $\chi_E = N$ and can be read directly from the binary symplectic description of the code. It therefore grows only linearly with the number of physical qubits, while the logical bond dimension attainable on $k$ encoded qubits is bounded by $2^{\lfloor k/2 \rfloor}$. Thus the compiled physical simulation stores exponentially less data than a direct logical simulation may require at the same depth. Numerically, random compositions drawn from the corresponding gate library generate logical states with volume-law entanglement entropy, non-trivial interaction distance from the Gaussian manifold~\cite{pachosSciPost,matos_emergence_2021,deger_persistent_2023}, and large magic measured by stabilizer R\'enyi entropy density~\cite{leone2022}. The physical bond dimension remains pinned at $\chi_E$ throughout.

The CCQC framework is also useful as a quantum hardware benchmarking tool. Direct fidelity estimation (DFE)~\cite{flammia2011} estimates the state fidelity between a target pure state and a noisy device output from a target-dependent distribution over Pauli observables, and its standard implementation requires the ability to compute ideal Pauli expectation values for the target. For CCQC outputs, these values are accessible through perfect Pauli sampling~\cite{lami2023} on the compiled MPS, followed by a Clifford pushback through the encoder, which produces both the sampled logical Pauli string and its signed
ideal value. The device executes the logical circuit and performs local Pauli readout on the requested observables. We demonstrate the protocol on the polar $[[16,11]]$
instance, where the encoded reference can be constructed exactly, allowing the ingredients of the protocol to be cross-checked against a direct logical evaluation. CCQCs thus allow us to stress-test the logical layer of an early fault-tolerant device with circuits that consist of long-range and non-Clifford gates.

Conceptually, our construction identifies both a structure in circuits that controls their classical simulation cost separately from gate class, depth, and entanglement content, and the existence of a physical representation in which the quantum dynamics solely contains onsite terms and classical relabeling. The encoder acts as the bridge between the two descriptions, in the same spirit in
which the Jordan--Wigner transformation maps a class of interacting spin models to free fermions or in which holographic codes~\cite{almheiri2015,pastawski2015} relate bulk and boundary degrees of freedom. The polar family studied here is one instance of the construction and the same procedure applies to any CSS code with an efficient encoder, a permutation-rich automorphism group, and a usable
transversal diagonal sector.

The remainder of the paper is organized as follows.  Section~\ref{sec:simulable} reviews the standard families of efficiently simulable circuits and the structural conditions that define each.  Section~\ref{sec:sim_prin} introduces the code-compiled quantum circuit (CCQC) construction and explains how code encoding maps a complicated logical circuit into onsite diagonal physical gates together with classical permutations.  Section~\ref{sec:gatefinder} describes how the logical gate library is generated from code automorphisms and transversal diagonal gates, and works through small-code examples.  Section~\ref{sec:polar_family} introduces the infinite polar CSS family used in the numerical study, and Section~\ref{sec:tn_sim} sets up the matrix-product-state simulation backend together with the classical relabeling mechanism. Section~\ref{sec-monomial} isolates the monomial subfamily handled by \textsc{PhasePoly.jl}, derives the corresponding Pauli-expectation sums, and states the precise caveat for level-$4$ phase terms. Section~\ref{sec:magic} defines the entanglement, non-Gaussianity, and magic diagnostics used to characterise the circuits.  Section~\ref{sec:numerical_evidence} reports the numerical results across polar instances, demonstrating that the standard resource measures grow with depth while the physical bond dimension stays pinned at the encoding cost.  Section~\ref{sec:dfe_application} develops the hardware-benchmarking application based on direct fidelity estimation and demonstrates it on a polar instance. Finally, in Sec.~\ref{sec:discussion}, we discuss the construction in the broader context of efficiently simulable scrambling dynamics and outline extensions to other codes and classical backends.

\section{Efficiently simulable circuits}
\label{sec:simulable}

Several families of quantum circuits admit efficient classical simulation, each defined by a particular set of restrictions on the allowed gates, input states, and measurement bases. We briefly recall the three families that are most directly relevant here. For each one we identify the structural assumptions that enable efficient simulation, and the sense in which our code-compiled circuits sit outside the regime in which those assumptions hold.

\subsection{Clifford circuits and the Gottesman--Knill theorem}

The Gottesman--Knill theorem~\cite{gottesman1998,dehaene2003,aaronson2004} provides an efficient classical simulation of any circuit built from Clifford gates, Pauli measurements, and classical feed-forward, provided the input is a stabilizer state such as $\ket{0}^{\otimes N}$. The standard representation is the stabilizer tableau: an $N$-qubit stabilizer state is encoded by $2N$ binary strings of length $2N+1$, and each Clifford gate updates this data by elementary row operations on a $2N \times (2N+1)$ binary matrix. The full state vector is never stored, and expectation values are read off directly from commutation relations with the stabilizer generators.

Efficient Clifford simulation rests on three restrictions. First, the gate set must remain Clifford: adding even a single non-Clifford gate such as $T$ together with Clifford gates and computational-basis measurements gives a universal gate set. Second, the initial state must be a stabilizer state, since Clifford circuits acting on non-stabilizer inputs, for example the magic state $\ket{T} = T\ket{+}$, already implement universal quantum computation through gate teleportation~\cite{gottesman_chuang1999,bravyi2005}. A line of work starting with Bravyi and Kitaev has extended these methods beyond the strict Clifford regime by decomposing the input or the non-Clifford gates themselves as superpositions of stabilizer states~\cite{bravyi2016,bravyi2018cliffordt}. The state $\ket{\psi}$ is written as $\ket{\psi}=\sum_{i=1}^{\chi_s}c_i\ket{\phi_i}$ with each $\ket{\phi_i}$ a stabilizer state, and a Clifford+$T$ circuit is then simulated by propagating the tableaus of every term in this decomposition in parallel. However, adding magic comes with an exponential cost. Third, measurements must be in Pauli bases, adaptive measurements in non-Pauli bases combined with Clifford gates again achieve universality~\cite{raussendorf2003}.

Our construction violates the first two of these restrictions simultaneously. The logical gate library contains $\CCZ$ and other non-Clifford diagonal gates, and the allowed input states need only admit an efficient initial MPS description rather than being stabilizer states. Either feature on its own already takes the family beyond the direct reach of Gottesman--Knill simulation, and crucially, neither the depth of the logical circuit nor the number of non-Clifford gates appears in the simulation cost, in contrast with the exponential cost of stabilizer-rank methods.

\subsection{Matchgate circuits and Gaussian (free-fermion) simulation}
\label{sec:matchgate}

Matchgates, introduced by Valiant~\cite{valiant2002}, are two-qubit gates of the form
\begin{equation}
G(A, B) = \begin{pmatrix} a_{00} & 0 & 0 & a_{01} \\ 0 & b_{00} & b_{01} & 0 \\ 0 & b_{10} & b_{11} & 0 \\ a_{10} & 0 & 0 & a_{11} \end{pmatrix},
\end{equation}
where $A = \begin{pmatrix} a_{00} & a_{01} \\ a_{10} & a_{11} \end{pmatrix}$ and $B = \begin{pmatrix} b_{00} & b_{01} \\ b_{10} & b_{11} \end{pmatrix}$ satisfy $\det A = \det B$. Under the Jordan--Wigner transformation, matchgate circuits on nearest-neighbour qubits correspond to free-fermion (Gaussian) evolution. The fermionic two-point correlation matrix evolves linearly, and Wick's theorem determines all higher-order correlations~\cite{terhal2002, jozsa_miyake_strelchuk_2015}.

The natural language for this description is in terms of Majorana operators $c_{2j-1} = \prod_{k<j} Z_k \cdot X_j$ and $c_{2j} = \prod_{k<j} Z_k \cdot Y_j$. A Gaussian state is fully characterised by the $2N \times 2N$ antisymmetric covariance matrix
\begin{equation}
\Gamma_{ab} = \frac{i}{2} \langle [c_a, c_b] \rangle.
\end{equation}
Any observable expressible through Wick's theorem reduces to a Pfaffian of a submatrix of $\Gamma$ and can therefore be evaluated in polynomial time.

As in the Clifford case, this efficiency comes with several structural restrictions. The gates must be matchgates, or equivalently the fermionic dynamics must be quadratic in Majorana operators thus non-matchgate two-qubit gates, such as $\SWAP$ between non-adjacent qubits, break the free-fermion structure. They must also act on nearest-neighbour qubits along a one-dimensional chain so that the Jordan--Wigner string remains intact, since long-range matchgates already promote the model to universal quantum computation~\cite{jozsa2008}. The initial state must be Gaussian, and the measurements must preserve the Gaussian character.

The matchgate framework can be pushed beyond its strict regime by allowing a controlled number of non-Gaussian resources, at an exponential cost in that number. Pure fermionic non-Gaussian states act as magic states for matchgate computations~\cite{hebenstreit_strelchuk_2019}. Mocherla, Lao, and Browne~\cite{mocherla2023} extend nearest-neighbour matchgate circuits with a handful of universality-enabling gates such as $\SWAP$, $\CZ$, or $\mathrm{CPhase}$, and show that single-qubit Pauli measurements on product inputs can still be simulated, but at a cost that grows exponentially in the number of non-matchgate gates. The analogous statement holds for non-Gaussian inputs: the runtime grows with the amount of non-Gaussianity, although there exists an intermediate regime in which certain structured non-Gaussian inputs remain efficiently approximable to additive error under free-fermion dynamics~\cite{Dias2024classicalsimulation, oh2026fermionic}.

Our family violates these restrictions on all three counts. The logical circuit contains $\CCZ$ and $\CCCZ$ terms, the induced logical connectivity is effectively all-to-all, and the allowed inputs need not be Gaussian, so the free-fermion description is absent both dynamically and at the level of the input state.

\subsection{Tensor network (MPS) simulation}

Matrix product state (MPS) simulation~\cite{vidal2003,schollwock2011} provides an efficient classical simulation of any quantum circuit whose entanglement remains bounded. For a fixed one-dimensional site ordering, an MPS with bond dimension $\chi$ represents any state whose Schmidt rank across every contiguous cut is at most $\chi$, and the simulation cost scales polynomially in $\chi$. The method is therefore efficient whenever $\chi$ stays polynomial in $N$ throughout the circuit. Unlike Clifford or matchgate simulation, there is no intrinsic restriction on the gate set or the input state; the only bottleneck is the growth of entanglement during the evolution.

This bottleneck has important consequences. Generic quantum circuits rapidly generate volume-law entanglement with $\chi \sim 2^{N/2}$, which makes tensor-network simulation exponentially costly. Our construction avoids this issue by confining all bond-dimension growth to the encoder. Every operation after encoding is either an onsite phase or a classically tracked permutation, so the physical MPS simulation remains efficient even when the logical dynamics look strongly scrambling.

\section{Code-compiled quantum circuits (CCQC)}
\label{sec:sim_prin}

We now introduce \emph{code-compiled quantum circuits} (CCQC), the central object of this paper. A CCQC is built in two layers. At the logical level it is a circuit on $k$ encoded qubits drawn from a structured gate set that includes nonlocal Clifford operations and non-Clifford diagonal gates such as controlled-$S^\dagger$, $\CCZ$ and $CCCZ$. At the physical level, it is the same circuit compiled through an error-correcting code, so that every layer after encoding reduces to single-qubit phases together with a classical relabelling of the qubits. The two descriptions are interchangeable, but their resource profiles differ sharply. The logical circuit can display several standard signatures usually associated with classical simulation hardness: it can generate volume-law entanglement, large nonstabilizerness, departure from the Gaussian manifold, and scrambling-like multipartite correlations. The compiled physical circuit, by contrast, is built from operations that an MPS simulator can absorb without any growth in bond dimension. This separation is what lets these circuits sit outside the Clifford, matchgate, and low-entanglement paradigms while remaining classically tractable.

We develop this construction in three steps. We first recall the stabilizer-code structure that defines the encoding map and the matched gate library. We then describe the compilation procedure that turns a logical circuit into its physical form. Finally, we show why the resulting physical circuit has low tensor-network simulation cost, with the entire bond-dimension budget paid once by the encoder.

\subsection{Stabilizer code background}

We consider a Calderbank--Shor--Steane (CSS) stabilizer code~\cite{gottesman_stabilizer_1997,steane1996,calderbank1996} with parameters $[[N,k,d]]$, which stores $k$ logical qubits in $N$ physical qubits by imposing separate commuting $X$- and $Z$-type parity checks. We write $\mathcal{S}$ for the stabilizer group and choose logical Pauli operators $\{\bar{X}_i,\bar{Z}_i\}_{i=1}^k$ to represent the encoded qubits. An encoding circuit $E$ is a Clifford unitary that maps
\[
E:\ket{\psi}\ket{0}^{\otimes(N-k)} \mapsto \ket{\bar{\psi}}\in \mathcal{H}_{\mathcal{S}},
\]
so that the logical state is embedded into the codespace of a larger physical register \cite{gottesman_surviving_nodate}. This notation distinguishes physical operators from logical operators and will be used throughout the paper.

The gate library used below is built from physical operations that preserve the codespace. One source is code automorphisms: physical Clifford unitaries $U$ which consist only of permutations, local Cliffords, and possible Pauli phase corrections, satisfying $U\mathcal{S}U^\dagger=\mathcal{S}$. Such symmetries induce well-defined logical Clifford gates~\cite{681315,grassl_leveraging_2013,hao_investigations_2021,sayginel2024}. In the code families studied here, the relevant automorphisms are usually qubit permutations, sometimes accompanied by single qubit Clifford operators, so simple qubit relabellings + single qubit operations at the physical level can correspond to dense logical circuits built from $\CNOT$, $H$, $S$, $X$, $Z$, and $\SWAP$ gates.

The second source is transversal diagonal operators that preserve the codespace~\cite{cui2017,Webster_2023,jain_high-distance_2024,eastin2009,koutsioumpas_smallest_2022}. Their induced logical action is diagonal in the logical $Z$ basis and is described by a phase polynomial; the explicit construction is given in Sec.~\ref{sec:gatefinder}.

For each code-preserving operation, we store both the physical circuit that acts on the physical qubits as well as its corresponding logical action circuit acting on the logical space. Specifically, for each automorphism and diagonal operator, we store: an automorphism pair $(A_r,\overline{A_r})$ or a diagonal pair $(D_r,\overline{D_r})$ where the bar denotes the logical action of the corresponding physical circuit.


\subsection{Efficient circuits via code compilation}

In our CCQC construction we do not use the code as a long-lived memory. Instead, we use the encoder as a compilation map: it carries the entangling cost, while the evolution after encoding is deliberately kept simple. For a fixed CSS code, we precompute the matched gate library in two steps. For automorphisms, following Ref.~\cite{sayginel2024}, we map the code to a binary linear description, identify the relevant symmetries, and recover the induced logical Clifford action together with any Pauli corrections. For diagonal gates, following Ref.~\cite{Webster_2023}, we solve the commutation constraints that determine which physical phase patterns preserve the codespace and what logical phase polynomial they induce.

We use the product input
\[
\ket{\psi_{\mathrm{in}}}=\ket{+}^{\otimes k}\ket{0}^{\otimes(N-k)}.
\]
Before encoding, the first $k$ qubits carry the logical input in $\ket{+}^{\otimes k}$, while the remaining $N-k$ auxiliary qubits are initialized in $\ket{0}$. After applying the encoder, we obtain the encoded plus state $\ket{\bar{+}}^{\otimes k}=E\ket{\psi_{\mathrm{in}}}$, which is the simultaneous $+1$ eigenstate of the full stabilizer group together with all logical $\bar{X}_i$. Because this encoded state is still a stabilizer state, we can determine its entanglement structure and bond dimension exactly from the encoded stabilizer tableau.

The automorphism pairs make up the Clifford sector of the library, and the diagonal pairs from the third level and above, make up the non-Clifford sector. Together they produce a logical gate set containing both Clifford and non-Clifford operations even though the compiled physical operations remain only permutations and onsite phase gates. Once these pairs have been computed for a given code, compiling a logical circuit amounts to selecting the desired matched operations and applying them in sequence.

A circuit specified by $R$ selected blocks $B_r$ has an encoded physical evolution
\begin{equation}\label{eq:circuit}
\ps{V}=\left(\prod_{r=1}^{R} \ps{B}_r\right)E,
\end{equation}
where the product is in the same time order as the selected blocks. This is the object simulated by the physical MPS backend. If one explicitly decodes at the end, then $E^\dagger \ps{V}$ acts on the logical register as
\begin{equation}
\bar{U}=\prod_{r=1}^{R} \bar{B_r}
\end{equation}
on the codespace, with the auxiliary qubits returned to their fixed syndrome sector in the ideal circuit. The efficient post-encoding simulation cost applies to $V_{\phys}$, while $E^\dagger$ is used only when we deliberately unencode for a different basis measurement. The logical circuit $\bar{U}$ can contain highly nonlocal Clifford layers and multi-qubit non-Clifford diagonal gates. Each block $B_r$ is a single automorphism or diagonal catalog entry, but it may expand into many elementary logical gates. $R$ therefore counts matched blocks, not the expanded logical gate count $L$.
The contrast between simple physical operations and complex logical action underlies the efficient simulability. 

\section{Automorphism and Diagonal gates}
\label{sec:gatefinder}

In this section, we make both mechanisms explicit on small codes: the $[[4,2,2]]$ code gives an automorphism-induced logical $\CNOT$, while the $[[8,3,2]]$ cube code gives a transversal logical $\CCZ$. We then apply the same two procedures to the scalable polar CSS family used in the numerical study.

Our focus is to use the symmetries of the error correcting code to allow for simple physical circuits that are easy to simulate, which yield complex logical circuits across many logical qubits. Transversal gates and code automorphisms are two classes of physical operations that exploit the symmetric structure of a code, and both have been of particular interest for performing low-overhead logic on error-correcting codes. We summarize both mechanisms here before turning to explicit small-code examples.

\subsection{Clifford hierarchy and diagonal gates}

The Clifford group on \(N\) qubits is the set of unitary operators that preserve the \(N\)-qubit Pauli group under conjugation:
\begin{equation}
\mathcal{C}_N := \left\{ U : U P U^\dagger \in \mathcal{P}_N
\;\text{for all}\; P \in \mathcal{P}_N \right\}.
\end{equation}
Equivalently, Clifford unitaries map Pauli operators to Pauli operators. Therefore, Clifford circuits map stabiliser states to stabiliser states.

The Clifford hierarchy is a nested sequence of gate sets defined recursively:
\begin{equation}\label{eq:hierarchy}
\mathcal{C}^{(1)} \equiv \mathcal{P}_N, \quad \mathcal{C}^{(t+1)} \equiv \{U : UPU^\dagger \in \mathcal{C}^{(t)} \;\forall\, P \in \mathcal{P}_N\}.
\end{equation}
The second level of the hierarchy, $\mathcal{C}^{(2)}$, is the Clifford group. These are unitary operations that map Pauli operators to Pauli operators, and are central to quantum error correction \cite{gottesman_stabilizer_1997, 681315}. Due to the Gottesman--Knill theorem \cite{gottesman1998, aaronson2004}, the action of Clifford operators  on stabilizer states can be tracked efficiently via tableau or symplectic representations~\cite{dehaene2003,aaronson2004}.
Although Clifford circuits alone are not computationally universal, access to any non-Clifford gate from the third level of the hierarchy promotes the Clifford group to a universal gate set, capable of approximating arbitrary quantum operations to any desired accuracy~\cite{Barenco_univ, bravyi2005}.
Examples of such gates include $T$ ($\mathrm{diag}(1, e^{i\pi/4})$), $\CS$
($\mathrm{diag}(1,1,1,i)$), and $\CCZ$
($\mathrm{diag}(1,1,1,1,1,1,1,-1)$). 

Following Refs.~\cite{cui2017,Webster_2023}, we focus on diagonal operators in the higher levels of the hierarchy, since these can be written as diagonal XP operators \cite{webster_xp_2022} and there are certain quantum error correcting codes yielding simple logical implementations of them \cite{jain_high-distance_2024}.

The fourth level of the hierarchy contains gates such as the controlled-$T$ gate
$\CT=\mathrm{diag}(1,1,1,e^{i\pi/4})$, the doubly controlled phase gate
$\CCS=\mathrm{diag}(1,1,1,1,1,1,1,i)$, and the four-qubit controlled phase
$\CCCZ=\mathrm{diag}(1,\ldots,1,-1)$. More generally, moving up the hierarchy
gives access to either finer diagonal rotations or controlled logic supported on
larger sets of qubits. 
Here we usually focus on the first four levels, the same framework extends to level $5$ and beyond, and some of the codes we study admit similarly simple physical implementations of these higher-level logical operations.

\subsection{Commutator method for finding transversal diagonal gates}

Before introducing the commutator method, we first fix the class of physical operations we will search over. These are the gates that remain simple at the physical level, yet can still induce nontrivial logical diagonal actions on the codespace.

\begin{definition*}[Transversal diagonal gate]
Let a code block consist of $N$ physical qubits. A physical gate is called \emph{transversal} if it ``factorizes'' across the qubits of the block, so that no elementary factor couples two physical qubits within the same block. In the single-block setting relevant here, a transversal diagonal gate is therefore any operator of the form
\[
D=\bigotimes_{i=1}^{N} d_i,
\]
where each $d_i$ is a one-qubit diagonal unitary.
\end{definition*}

Such gates are natural from the fault-tolerant perspective and are especially convenient for our simulation setting, since they act onsite and therefore do not by themselves increase the MPS bond dimension. 

The Eastin--Knill theorem~\cite{eastin2009} shows that no quantum error-detecting code can implement a universal gate set using only transversal gates.
For our purposes, this means that the resulting library is necessarily structured rather than universal, but that is exactly the regime we need: a gate set rich enough to generate nontrivial entanglement and magic while still admitting a controlled physical implementation.

We now describe how to determine which of these transversal diagonal phase layers preserve the code space and what logical phase polynomial they induce.

Following Refs.~\cite{cui2017,Webster_2023}, we represent transversal diagonal operators at level $t$ of the Clifford hierarchy as vectors modulo $M:=2^t$ and search for those that preserve the code space. 
For a CSS code described by its $X$-check matrix $S_X$ and $X$-logical matrix $L_X$, we first find the diagonal logical identities which fix all vectors in the codespace.
The logical identities can be found by calculating a kernel over the ring $\mathbb{Z}_M$. 
A transversal diagonal operator acts as a logical operator on a CSS code if and only if its group commutator with each of the X-checks is a diagonal logical identity. 
Such operators can be identified efficiently by solving a series of linear constraints over $\mathbb{Z}_M$.
We compute the induced logical phase polynomial of the logical operator by calculating its action under conjugation on the logical X basis.

At the level-$3$ search, the resulting logical phase polynomial can contain degree-$1$ terms corresponding to logical $Z_i$ gates, degree-$2$ terms corresponding to controlled-phase gates such as $\CZ_{i,j}$ or $\CS^\dagger_{i,j}$, and degree-$3$ terms corresponding to $\CCZ_{i,j,k}$. This is the same diagonal-gate language used in the worked examples below.

\subsection{Automorphisms of a stabilizer code}

We now turn to the second class of physical circuits that yield easy to simulate logical gates, which are based on code automorphisms. These are the physical Clifford symmetries that preserve the stabilizer structure and therefore act as logical Clifford gates on the encoded qubits.

\begin{definition*}[Automorphism of a stabilizer code]
Let $\mathcal{S}$ be the stabilizer group of a code. A physical Clifford unitary $U$ is called an \emph{automorphism} of the code if it consists only of permutations, local Cliffords, and possible Pauli phase corrections, and satisfies:
\[
U \mathcal{S} U^\dagger = \mathcal{S}.
\]
In this case, $U$ preserves the codespace and induces a well-defined logical Clifford operation on the encoded qubits. The set of all such automorphisms forms the automorphism group, denoted
\[
\Aut(\mathcal{S})
\]
\end{definition*}

In this work, we identify such symmetries by mapping the stabilizer code to a binary linear code, computing a corresponding graph automorphism group, and then recovering the induced logical action together with any required Pauli corrections.

We find code automorphisms by constructing an auxiliary graph encoding the stabilizer structure, the so-called Tanner graph~\cite{TannerG}, augmented with additional edges and using a graph-automorphism algorithm (such as Bliss~\cite{junttila2007} or Nauty~\cite{MCKAY201494}), following the methods introduced in Ref.~\cite{sayginel2024}. We implement this search in the following steps:
\begin{enumerate}
\item Starting from the stabilizer generators $[H_X | H_Z]$ construct the three block form by finding $H := [H_X | H_Z| H_X \oplus H_Z]$.
\item Construct the Tanner graph $G$ of the binary linear code with parity check matrix $H$ by adding a vertex for each column and row. Vertices corresponding to rows are colored blue, and vertices corresponding to columns are colored red.  
\item Edges encode which bits participate in which check, i.e., we draw an edge between column node $c_i$ and check node $v_j$ iff $H_{j,i}=1$.
\item Add edges between bits which correspond to the same qubit -- this limits column permutations to those which corresponds to qubit permutations and single qubit Clifford gates.
\item Find the automorphism group $\Aut(G)$ using a graph automorphism package.
\item Extract the induced logical Clifford by tracking the action on the logical operators and adding the required Pauli corrections following ~\cite{sayginel2024}.
\end{enumerate}

This method allows us to find automorphisms of large codes in a short amount of time and study their logical gates efficiently. Note that as discussed in ~\cite{sayginel2024}, this does not necessarily yield the full automorphism group of the code, but rather a subgroup, providing a tradeoff between speed and accuracy, i.e., $\Aut(G) \subseteq  \Aut(\mathcal{S})$. For the examples considered here, however, this quickly yields a big enough group to showcase our methods.

\subsection{Worked examples}
\label{sec:worked_examples}
With these logical operators in hand, we now illustrate them on small codes where the induced logical action can be written out explicitly. We first use the $[[4,2,2]]$ code, a four-qubit CSS code encoding two logical qubits, to show how a simple physical permutation becomes an entangling logical Clifford. We then use the $[[8,3,2]]$ cube code, an eight-qubit three-dimensional colour-code instance encoding three logical qubits, to show how an onsite diagonal phase layer becomes a logical $\CCZ$. These examples contain the main ingredients we use for the larger polar-family construction.

\subsubsection{\texorpdfstring{Automorphism example: the $[[4,2,2]]$ code}{Automorphism example: the [[4,2,2]] code}}
\label{sec:worked_example}

We begin with the $[[4,2,2]]$ code, a four-qubit CSS code encoding two logical qubits with stabilizers
\[
\mathcal{S}=\langle X_1X_2X_3X_4,\;Z_1Z_2Z_3Z_4\rangle.
\]
We choose the logical basis
\begin{align}
\bar{X}_1 &= X_1X_2, & \bar{Z}_1 &= Z_1Z_3, \notag\\
\bar{X}_2 &= X_1X_3, & \bar{Z}_2 &= Z_1Z_2.
\end{align}
Figure~\ref{fig:422_square} shows the square geometry and the bipartition underlying this choice of logical basis.
Because the stabilizers are fully symmetric, any transposition of physical qubits preserves the code space. In the language of Ref.~\cite{sayginel2024}, every such permutation is a code automorphism, and its logical action is recovered by tracking the logical Paulis. For the physical relabeling $\SWAP_{1,2}$ one finds
\[
\bar{X}_1 \mapsto \bar{X}_1,\qquad
\bar{Z}_1 \mapsto Z_2Z_3=\bar{Z}_1\bar{Z}_2,
\]
\[
\bar{X}_2 \mapsto X_2X_3=\bar{X}_1\bar{X}_2,\qquad
\bar{Z}_2 \mapsto \bar{Z}_2.
\]
This is exactly the Heisenberg action of $\overline{\CNOT}{_{2,1}}$~\cite{gottesman1998}, so a single physical qubit swap induces an entangling logical Clifford gate:
\[
U_{\phys}=\SWAP_{1,2}\qquad\Longrightarrow\qquad \bar{U}=\overline{\CNOT}{_{2,1}}.
\]
This is the same mechanism used later in the polar family: the physical operation is only a relabeling of qubits, yet the induced logical Clifford can be entangling and, for larger codes, highly nonlocal.

\begin{center}
\begin{tikzpicture}[scale=1.4]

\tikzstyle{qubitA}=[circle,draw,fill=blue!60,minimum size=7pt]
\tikzstyle{qubitB}=[circle,draw,fill=red!60,minimum size=7pt]

\node[qubitA,label=above:$q_1$] (q1) at (0,1) {};
\node[qubitB,label=above:$q_2$] (q2) at (2,1) {};
\node[qubitA,label=below:$q_3$] (q3) at (2,-1) {};
\node[qubitB,label=below:$q_4$] (q4) at (0,-1) {};

\draw[thick] (q1) -- (q2) -- (q3) -- (q4) -- (q1);

\end{tikzpicture}
\end{center}
\inlinefigcaption{fig:422_square}{
Geometric representation of the $[[4,2,2]]$ code. The two diagonals (blue and red vertices) provide a convenient way to organise the logical basis used in the text, e.g.\ $\bar Z_1=Z_1Z_3$ is supported on the blue diagonal.
}

\subsubsection{\texorpdfstring{Diagonal example: the $[[8,3,2]]$ cube code}{Diagonal example: the [[8,3,2]] cube code}}
\label{sec:cube_example}

We next use the $[[8,3,2]]$ cube code~\cite{bombin2006,bombin2015,vasmer2019,menendez2024}, which is a standard small code example of a transversal non-Clifford logical gate. Figure~\ref{fig:832_cube} shows the cube labeling and the signed $T/T^\dagger$ pattern used below. We use the generating set
\[
\begin{aligned}
\mathcal{S}=\langle &X_1X_2X_3X_4X_5X_6X_7X_8,\;
Z_1Z_2Z_3Z_4,\;
Z_5Z_6Z_7Z_8, \\
&Z_1Z_2Z_5Z_6,\;
Z_1Z_3Z_5Z_7\rangle,
\end{aligned}
\]
and we choose the logical Pauli basis
\begin{align}
\bar{X}_1 &= X_1X_2X_3X_4, & \bar{Z}_1 &= Z_1Z_5, \notag\\
\bar{X}_2 &= X_1X_2X_5X_6, & \bar{Z}_2 &= Z_1Z_3, \notag\\
\bar{X}_3 &= X_1X_3X_5X_7, & \bar{Z}_3 &= Z_1Z_2.
\end{align}
The three logical $X$ supports are the three faces incident on vertex~$1$, so
\[
\left|\operatorname{supp}(\bar{X}_1)\cap \operatorname{supp}(\bar{X}_2)\cap \operatorname{supp}(\bar{X}_3)\right|=1.
\]
This single-point triple overlap is the geometric reason a cubic logical phase can appear: the signed onsite phases can cancel all linear and quadratic contributions, while the single common vertex leaves a surviving three-body term. 

Consider the onsite phase layer
\[
U_{\phys} = T_1 T_2^\dagger T_3^\dagger T_4 T_5^\dagger T_6 T_7 T_8^\dagger.
\]
At level $3$, using the notation from \cite{Webster_2023} we write this as the diagonal XP operator:
\[
U_{\phys} = XP_8(0|\mathbf{0}|\mathbf{z}),
\qquad
\mathbf{z}=(1,7,7,1,7,1,1,7),
\]
where $1$ denotes $T$ and $7\equiv -1 \pmod 8$ denotes $T^\dagger$. For a non-Clifford diagonal gate, the relevant condition is preservation of the stabilizer-defined code space rather than Pauli-to-Pauli conjugation of every generator. Here that check is transparent. Because $U_{\phys}$ is diagonal, it commutes with all four $Z$-type stabilizers. In the canonical codeword notation,
\begin{align}
\text{with }\mathbf{x}&=(x_1,x_2,x_3), \notag\\
\ket{\bar{x}_1\bar{x}_2\bar{x}_3}
&= \frac{1}{\sqrt{2}}\sum_{u\in\mathbb{Z}_2}\ket{uS_X + \mathbf{x}L_X} \notag\\
&= \frac{\ket{c(\mathbf{x})}+\ket{\mathbf{1}\oplus c(\mathbf{x})}}{\sqrt{2}}, \notag\\
c(\mathbf{x})
&= \mathbf{x}L_X \notag\\
&= \bigl(0, x_3, x_2, x_2\oplus x_3, \notag\\
&\qquad x_1, x_1\oplus x_3, x_1\oplus x_2, x_1\oplus x_2\oplus x_3\bigr),
\end{align}
where $\mathbf{1}$ is the all-ones string.
Here $L_X$ denotes the representatives $(00001111;\,00110011;\,01010101)$, which differ from the logical basis above only by multiplication with the all-$X$ stabilizer. This choice makes the codeword expansion compact.
The $Z$ stabilizers pair each computational basis string with its complement, and the all-$X$ stabilizer fixes the symmetric superposition within each pair. Writing $\omega_8=e^{i\pi/8}$, the physical phase layer acts as
\begin{align}
U_{\phys}\ket{\mathbf{e}}
&=\omega_8^{2\mathbf{e}\cdot\mathbf{z}}\ket{\mathbf{e}}, \notag\\
2(\mathbf{1}\cdot\mathbf{z})
&\equiv 0 \pmod{16}, \notag\\
2c(\mathbf{x})\cdot\mathbf{z}
&=2(\mathbf{1}\oplus c(\mathbf{x}))\cdot\mathbf{z} \notag\\
&=8x_1x_2x_3 \pmod{16}.
\end{align}
Hence
\[
\bar{U}\ket{\bar{x}_1\bar{x}_2\bar{x}_3}=(-1)^{x_1x_2x_3}\ket{\bar{x}_1\bar{x}_2\bar{x}_3},
\]
so the code space is preserved and only $\ket{\bar{1}\bar{1}\bar{1}}$ acquires a minus sign. In the controlled-phase language of Ref.~\cite{Webster_2023}, all weight-$1$ and weight-$2$ coefficients vanish and the only surviving logical term is $CP_8(8,\mathbf{111})=\overline\CCZ_{1,2,3}$. The same cubic phase polynomial reproduces the logical Heisenberg action~\cite{gottesman1998}:
\begin{align}
\bar{X}_1 &\mapsto \bar{X}_1\,\overline\CZ_{2,3}, \notag\\
\bar{X}_2 &\mapsto \bar{X}_2\,\overline\CZ_{1,3}, \notag\\
\bar{X}_3 &\mapsto \bar{X}_3\,\overline\CZ_{1,2},
\end{align}
which is the defining action of $\CCZ_{1,2,3}$. Hence
\[
U_{\phys}\qquad\Longrightarrow\qquad \bar{U}=\overline\CCZ_{1,2,3}.
\]
This is the small code prototype of the non-Clifford sector used later: in the polar family, the same diagonal search returns matched onsite phase layers whose logical action is a phase polynomial containing logical $T$, controlled-$S^\dagger$, and $\CCZ$ terms on larger sets of encoded qubits.

\begin{center}
\begin{tikzpicture}[scale=1.4]

\tikzstyle{qubitT}=[circle,draw,fill=green!60,minimum size=7pt]
\tikzstyle{qubitTd}=[circle,draw,fill=orange!70,minimum size=7pt]

\node[qubitT,label=below left:$q_1$] (a) at (0,0) {};
\node[qubitTd,label=below:$q_2$] (b) at (2,0) {};
\node[qubitT,label=above:$q_4$] (c) at (2,2) {};
\node[qubitTd,label=above left:$q_3$] (d) at (0,2) {};

\node[qubitTd,label=left:$q_5$] (e) at (0.7,0.7) {};
\node[qubitT,label=below:$q_6$] (f) at (2.7,0.7) {};
\node[qubitTd,label=right:$q_8$] (g) at (2.7,2.7) {};
\node[qubitT,label=above:$q_7$] (h) at (0.7,2.7) {};

\draw[thick] (a)--(b)--(c)--(d)--(a);
\draw[thick] (e)--(f)--(g)--(h)--(e);
\draw[thick] (a)--(e);
\draw[thick] (b)--(f);
\draw[thick] (c)--(g);
\draw[thick] (d)--(h);

\end{tikzpicture}
\end{center}
\inlinefigcaption{fig:832_cube}{Cube representation of the $[[8,3,2]]$ code. The qubit labels match the stabilizers and logical operators written above. Green vertices carry $T$ and orange vertices carry $T^\dagger$ in the transversal phase layer.}

\section{Polar CSS family}
\label{sec:polar_family}
We now turn from the small worked examples to the scalable family used in our numerical study. The polar CSS family combines the same two gate-generation mechanisms in a high-rate construction whose encoder, automorphism search, and diagonal-gate search remain tractable as the blocklength grows. It arises from channel polarization~\cite{7208851,PhysRevLett.113.030501,10619465,arikan2009channel} and uses the same $F^{\otimes l}$ row structure that underlies Reed--Muller constructions~\cite{PhysRevA.54.4741,PhysRevA.86.052329,PhysRevLett.120.050504}.

\subsection{Hadamard construction and row selection}

The construction starts from the $2 \times 2$ matrix: 
\begin{equation}
F = \begin{pmatrix} 1 & 0 \\ 1 & 1 \end{pmatrix},
\end{equation}
and for blocklength $N=2^l$ we define
\begin{equation}
G_N = F^{\otimes l},
\end{equation}
where $F^{\otimes l} := \bigotimes_{i=1}^{l} F
$ ~\cite{arikan2009channel,renes2012efficient}. We therefore specify the CSS family by selecting structured rows from this matrix based on different selection criteria.

In our construction, we pick a low weight row (such as the unique weight 1 row) as the single $X$ check, while the $k$ highest-weight rows provide the logical $X$ representatives ~\cite{renes2012efficient,wilde2013polar}. Taking the kernel of the matrix whose rows are the $X$ checks and logicals, gives us the $Z$ checks of the CSS code, and from there we can find the remaining logical $Z$ representatives. The checks form the parity check matrix of the code, which is enough to identify an encoding circuit.

As an explicit example, consider the $l{=}3$ instance $[[8,4]]$. Indexing the rows of $G_8=F^{\otimes 3}$ by $u\in\{0,1\}^3$, row $u$ is the indicator vector of the subcube $\{x : x \subseteq u\}$ and has weight $2^{|u|}$. The weight-one row ($u=000$) is the single $X$ check, the four rows of weight at least four ($u\in\{011,101,110,111\}$) are the logical $X$ representatives, and the kernel of the matrix yields the three $Z$ checks:
\begin{equation}
\label{eq:l3}
\begin{aligned}
S_X &= \begin{pmatrix} 1&0&0&0&0&0&0&0 \end{pmatrix},\\[2pt]
S_Z &= \begin{pmatrix}
0&1&0&1&0&1&0&1\\
0&0&1&1&0&0&1&1\\
0&0&0&0&1&1&1&1
\end{pmatrix},\\[2pt]
L_X &= \begin{pmatrix}
1&1&1&1&0&0&0&0\\
1&1&0&0&1&1&0&0\\
1&0&1&0&1&0&1&0\\
1&1&1&1&1&1&1&1
\end{pmatrix},
\end{aligned}
\end{equation}
with paired logical $Z$ representatives $\bar Z_1 = Z_4Z_8$, $\bar Z_2 = Z_6Z_8$, $\bar Z_3 = Z_7Z_8$, and $\bar Z_4 = Z_8$. Note that the three $Z$ checks are the bitwise complements of the three weight-four logical rows.

\subsection{Parameters, encoder complexity, and gate library}

This choice produces a family with parameters
\begin{equation}\label{eq:params}
N = 2^l,
\qquad
k = 2^l - l - 1,
\end{equation}
and rate $k/N = 1 - (l+1)/2^l \to 1$ as $l \to \infty$. The distance of the induced CSS code is $1$, as there is only one $X$ check on a single qubit, but distance is not a critical consideration here. We use the code as a compilation gadget, so the important features are high rate, low encoder complexity, and the structured symmetries needed to generate a rich logical gate library. In the numerical study we focus on the $l=4$ and $l=5$ instances, namely $[[16,11]]$ and $[[32,26]]$.

The same recursive Kronecker-product structure also makes the encoder efficient. The map $x = u G_N$ can be implemented by a binary Fast Hadamard Transform using $\frac{N}{2}\log_2 N$ CNOT gates, i.e. $\Theta(N \log N)$ two-qubit operations~\cite{arikan2009channel}. This is the only stage of the compiled simulation that can increase the MPS bond dimension. The same row structure simultaneously generates the gate library: automorphism searches over physical permutations produce matched logical Clifford gates, while the level-3 and level-4 commutator searches produce logical operators whose actions include products of diagonal phase layers containing single-, two-, and three-body logical phases. We focus on levels $t=3$ and $t=4$ because they already yield a non-Clifford library rich enough for the benchmarks in our study.Sec.~\ref{sec:numerical_evidence}; higher-level diagonal searches still leave the post-encoding cost bound unchanged, since after compilation they remain onsite diagonal operations. What matters is that the search returns many nontrivial matched permutations, enough to furnish a nonlocal logical Clifford sector while keeping the physical cost fixed by the encoder. For the $l=5$ instance we export the full matched catalog and sample from it in Sec.~\ref{sec:numerical_evidence}. 

The same construction extends to higher levels of the Clifford hierarchy. At level $t=4$ the diagonal search returns matched pairs whose physical action is again a layer of single-qubit phases (now of the form $\sqrt{T}=P(\pi/8)$, $\sqrt{T}^\dagger$, and finer rotations) and whose induced logical action contains gates such as $\CT$, $\CCS$, and $\CCCZ$ on small sets of logical qubits. The key point is that the post-encoding cost bound only requires the compiled physical layers to be onsite, so moving from level $3$ to level $4$ does not change the encoding-set bond dimension and only the catalog of available logical gates becomes richer. Therefore, throughout the paper the circuit library is a mixed catalog of matched pairs: level-$3$ and level-$4$ entries are stored and sampled in the same way and use the same polar encoder. The $[[16,11]]$ instance is worked out explicitly in Appendix~\ref{app:t4_examples}.

The automorphism side of the library scales similarly. For the polar instances studied here, the qubit-permutation subgroup recovered by the Tanner-graph search is already large at $l=5$, producing many independent matched logical Clifford operations. A handful of automorphism layers is enough to spread support across all $k=26$ logical qubits, so the matched library is rich enough to drive genuinely many-body logical dynamics even though every compiled physical layer is just a permutation followed by single qubit gates. We give a table of the automorphism groups computed and highest level of the diagonal gate hierarchy for which the physical single qubit gate + permutation circuits give rise to multi-qubit logical operations in Table \ref{tab:aut_diag_levels_compact}.


\begin{table}[htbp]
\centering
\setlength{\tabcolsep}{4pt}
\begin{tabular}{@{}rrrrc@{}}
\toprule
$l$ & $N$ & $k$ & $|\mathrm{Aut}|$ & $t$ \\
\midrule
2 & 4  & 1  & $2^{5}\!\cdot\!3$ & 0 \\
3 & 8  & 4  & $2^{11}\!\cdot\!3\!\cdot\!7$ & 3 \\
4 & 16 & 11 & $2^{22}\!\cdot\!3^{2}\!\cdot\!5\!\cdot\!7$ & $\geq 6$ \\
5 & 32 & 26 & $2^{42}\!\cdot\!3^{2}\!\cdot\!5\!\cdot\!7\!\cdot\!31$ & $\geq 6$ \\
6 & 64 & 57 & $2^{79}\!\cdot\!3^{4}\!\cdot\!5\!\cdot\!7^{2}\!\cdot\!31$ &  $\geq 6$ \\
\bottomrule
\end{tabular}
\caption{Orders of the code-compiled quantum circuits for the polar codes constructed. The first three columns list the code parameters obtained from the construction of Eq \ref{eq:params}. The fourth column lists the order of the automorphism groups computed whereas the last column lists the highest level of Clifford hierarchy for which physical single-qubit gates on the data qubits of the code give rise to multi-qubit logical actions.}
\label{tab:aut_diag_levels_compact}
\end{table}

\section{Simulating physical circuits using Matrix Product States}
\label{sec:tn_sim}

We now explain why the compiled physical circuit of Eq.~\eqref{eq:circuit} is cheap to simulate as an MPS. The main idea is as follows: After encoding, every layer is a product of single-qubit Clifford and phase operators, together with a permutation of qubit labels, and neither operation can grow the bond dimension of the stored state.

We place the $N$ physical qubits on a fixed MPS chain and keep a small classical lookup table $\mu$ that records, for each physical qubit $i$, which site of the chain currently holds it. A diagonal physical layer $D^{\phys}_r=\bigotimes_i d_i^{(r)}$ is then applied site by site; because each $d_i^{(r)}$ acts on a single qubit, it commutes with the Schmidt decomposition at every cut and the bond dimension is unchanged. The permutation layer within $A^{\phys}_r$ is treated as a passive qubit-label update rather than as an active SWAP network on the stored MPS. Thus we simply update $\mu$ in $O(N)$ time so that any later onsite gate acting on physical qubit $i$ is dispatched to chain site $\mu(i)$. The stored tensors are never touched, so the bond dimension is again unchanged. Iterating over all $R$ pairs of circuits, the only part of the simulation that can increase the bond dimension is the encoder $E$ itself.

Determining the simulation complexity is straightforward. Let $\chi_E$ be the largest bond dimension reached anywhere inside the encoder. The encoder contains $|E|$ two-qubit gates, each of which costs $O(\chi_E^3)$ in the standard MPS update~\cite{vidal2003,schollwock2011}. Every matched block afterwards adds only $O(N)$ single-qubit updates of cost $O(\chi_E^2)$ each. Summing over the $R$ matched blocks gives the total simulation cost
\begin{equation}
\mathrm{cost}_{\mathrm{tot}}(R)=\mathcal{O}\!\left(|E|\,\chi_E^3 + R N \chi_E^2\right),
\label{eq:cost_bound}
\end{equation}
which is linear in the number of matched blocks $R$.

For the stabilizer reference state, the final encoded-state Schmidt ranks can be computed exactly from the check matrix using linear algebra. Since the peak along any encoder path is at least the final value, this is a lower bound on $\chi_E$. For the polar encoder ordering used in our simulations the bound is tight: the observed peak during encoding equals the final encoded-state value. For non-stabilizer inputs, $\chi_E$ should be understood as the actual peak reached along the chosen encoding path.

The required bond dimension can be calculated directly from the stabilizer generators~\cite{fattal2004,nahum2017}. Represent each Pauli generator by a binary row $(x|z)$: on each qubit, $x=1$ for an $X$ or $Y$, and $z=1$ for a $Z$ or $Y$. Stacking $N$ independent generators gives a matrix $G\in\Fbb_2^{N\times 2N}$. For a cut $A|B$, let
\[
G_B\in\Fbb_2^{N\times 2|B|}
\]
be the part of $G$ that describes the action on the qubits in $B$. Then the Schmidt rank across the cut is
\begin{equation}
\chi(A|B)=2^{|A|-N+\operatorname{rank}_{\Fbb_2}(G_B)},
\label{eq:chi_from_rank}
\end{equation}
and we define
\begin{equation}\label{eq:chiE_def}
\chi_E := \max_{A|B}\chi(A|B).
\end{equation}

Here is a direct derivation. Let $\operatorname{Stab}(\psi)$ denote the full stabilizer group of the state. A binary vector $c\in\Fbb_2^N$ selects a product of its generators. The combinations that act trivially on $B$ form the left kernel
\[
K_A:=\{c\in\Fbb_2^N:cG_B=0\}.
\]
By rank--nullity, the number of independent such combinations is
\[
d_A:=\dim_{\Fbb_2}(K_A)=N-\operatorname{rank}_{\Fbb_2}(G_B).
\]
These combinations generate the subgroup of stabilizers supported only on $A$,
\[
\operatorname{Stab}_A(\psi)
:=\{g\in\operatorname{Stab}(\psi):g=g_A\otimes I_B\}.
\]
Because the original generators are independent, this subgroup contains
\[
|\operatorname{Stab}_A(\psi)|=2^{d_A}
\]
distinct stabilizers.
The stabilizer-state density matrix is the uniform group average. When we trace out $B$, only the elements of $\operatorname{Stab}_A(\psi)$ survive:
\[
\rho_A=\Tr_B(\rho)
=2^{-|A|}\sum_{g\in\operatorname{Stab}_A(\psi)}g_A.
\]
This is proportional to the projector onto their common $+1$ eigenspace. Each of the $d_A$ independent stabilizers halves the dimension of that eigenspace, so
\[
\operatorname{rank}(\rho_A)
=\frac{2^{|A|}}{|\operatorname{Stab}_A(\psi)|}
=2^{|A|-d_A}
=2^{|A|-N+\operatorname{rank}_{\Fbb_2}(G_B)}.
\]
For a pure state, $\operatorname{rank}(\rho_A)$ is the Schmidt rank across $A|B$, which proves Eq.~\eqref{eq:chi_from_rank}.

It is useful to verify the formula on two elementary two-qubit states. For the Bell state, generated by $\{XX,ZZ\}$, consider the cut $1|2$ with $A=\{1\}$ and $B=\{2\}$. Restricting to subsystem $B$ gives
\[
G_B=
\begin{pmatrix}
1 & 0\\
0 & 1
\end{pmatrix},
\]
so $\operatorname{rank}(G_B)=2$. Since $N=2$ and $|A|=1$, Eq.~\eqref{eq:chi_from_rank} gives
\[
\chi(A|B)=2^{1-2+2}=2,
\]
as expected. For the product state $\ket{+}\ket{0}$, generated by $\{XI,IZ\}$, the same cut gives $\operatorname{rank}(G_B)=1$, and therefore
\[
\chi(A|B)=2^{1-2+1}=1,
\]
again as expected.

For an $[[N,k]]$ CSS code with $X$-check matrix $S_X$, $Z$-check matrix $S_Z$, and logical-$X$ matrix $L_X$, the encoded plus state $\ket{\bar{+}}^{\otimes k}$ is stabilized by every $X$-check, every $Z$-check, and every logical $\bar X_i$. In symplectic form these generators appear as the rows of the block matrix
\begin{equation}
G = \begin{pmatrix} S_X & 0 \\ 0 & S_Z \\ L_X & 0 \end{pmatrix}
\in \Fbb_2^{N \times 2N},
\label{eq:G_from_code}
\end{equation}
which has $m_X+m_Z+k=N$ rows in total, and these rows are independent because the $\bar X_i$ lie outside the stabilizer group by definition. Substituting Eq.~\eqref{eq:G_from_code} into Eq.~\eqref{eq:chi_from_rank} therefore reads $\chi_E$ off directly from the code matrices, with no need to simulate the encoder. In the polar family studied below, this evaluation gives $\chi_E=2^l=N$ for $l\ge 3$.

For comparison, the logical Schmidt rank obeys a general upper bound. Across a bipartition with $|A|=a$ and $|B|=k-a$, the Schmidt rank cannot exceed the smaller Hilbert-space dimension, so
\[
\chi(A|B)\le \min(2^a,2^{k-a}) = 2^{\min(a,k-a)}.
\]
Maximising over $a$ gives
\begin{equation}
\max_{A|B}\chi(A|B)\le 2^{\lfloor k/2\rfloor},
\label{eq:logical_chi_bound}
\end{equation}
attained by any state maximally entangled across a balanced cut (for example a tensor product of $\lfloor k/2\rfloor$ Bell pairs, together with one decoupled qubit when $k$ is odd). This logical upper bound is exponential in the number of logical qubits. By contrast, the physical bond dimension of the encoded reference state is fixed by the stabilizer structure of the code through Eq.~\eqref{eq:chiE_def}. Random compositions from the logical gate library, especially those involving many nonlocal gates, are expected to generate strongly scrambling logical dynamics~\cite{nahum2017,hayden2007,sekino2008}: the all-to-all Clifford connectivity induced by automorphisms, together with non-Clifford $\CCZ$ interactions, can drive entanglement toward volume-law scaling and push $\chi_{\logi}$ toward the bound in Eq.~\eqref{eq:logical_chi_bound}.

We emphasize that the simulation itself is not restricted to the stabilizer reference state.
The Gottesman--Knill theorem requires stabilizer-state inputs; matchgate simulation requires Gaussian inputs with nearest-neighbor connectivity; IQP circuits conventionally start from $\ket{+}^{\otimes N}$.
Our construction, by contrast, accepts any input representable as an MPS of moderate bond dimension $\chi_0$: product states, entangled states, or states with high magic or non-Gaussianity. The controlling factor is the peak bond dimension $\chi_E$ reached after encoding, which depends on $\chi_0$ and the code but not on the number of matched blocks $R$.

We note that alternative preparation strategies can also be explored. Rather than reaching the codespace by a fixed unitary encoder $E$, one could start from a simple product state and measure the stabilizer checks of the code, applying the usual Pauli-frame corrections conditioned on the outcomes. In an MPS simulation this remains a well-defined state preparation. Such measurements may reduce the practical cost of reaching the codespace for some codes, especially when the measured checks are local or otherwise cheap in the chosen tensor-network ordering. They do not, however, make an arbitrarily difficult input state free: the controlling quantity for the MPS backend is the largest bond dimension $\chi_E$ reached during preparation, regardless of whether that preparation is unitary or measurement based.

With the stabilizer-rank formulas in hand, we can now present a concrete bond-dimension calculation for stabilizer initial states in the polar family. For the polar $l{=}3$ code $[[8,4]]$, the $X$- and $Z$-check matrices and logical $X$ representative are given in Eq.~\eqref{eq:l3}. Building $G$ from the CSS block form of Eq.~\eqref{eq:G_from_code} at each cut gives $\chi = (1,2,4,4,8,4,2)$ hence $\chi_E = 8$. This matches the result obtained by propagating through all four encoding gates.

\begin{figure}[t!]
\algstart{Code-compiled MPS simulation}{alg:simulation}
\begin{algorithmic}[1]
\State \textbf{Input:} a CSS code $[[N,k,d]]$ with encoding circuit $E$, a library of automorphism pairs $\{(A^{\phys}_i,\overline{A}_i)\}$ and diagonal pairs $\{(D^{\phys}_j,\overline{D}_j)\}$, a logical circuit specified by $R$ matched blocks, and an initial state $\ket{\psi_0}$ written as an MPS
\vspace{2pt}
\Statex \textbf{Phase 1: Encoding}
\State Prepare an $N$-qubit MPS by placing $\ket{\psi_0}$ on the $k$ logical qubits and $\ket{0}$ on the remaining $N-k$ auxiliary qubits
\State Apply the encoding circuit $E$ to obtain the encoded MPS
\vspace{2pt}
\Statex \textbf{Phase 2: Logical evolution with on-the-fly relabeling}
\State Initialise a qubit-to-site map $\mu(i)\gets i$ for $i=1,\dots,N$
\For{$r=1,\dots,R$}
    \State Update the qubit map $\mu$ according to the physical automorphism $A^{\phys}_r$
\For{each qubit $i \in \mathrm{supp}(D^{\phys}_r) \cup \mathrm{supp}(A^{\phys}_r)$}
    \State Apply the local diagonal gate from $D^{\phys}_r$, if present, to MPS site $\mu(i)$
    \State Apply the on-site Clifford gate from $A^{\phys}_r$, if present, to MPS site $\mu(i)$
\EndFor
\EndFor
\vspace{2pt}
\Statex \textbf{Phase 3: Observables}
\State Compute expectation values and correlation functions directly from the MPS
\State For $Z$-basis sampling, sample directly from the encoded MPS
\State Use the map $\mu$ to identify the physical sites corresponding to the desired logical operators
\end{algorithmic}
\algfinish
\end{figure}

Applying the same procedure for polar instances $l=2,\ldots,7$ gives:
\begin{center}
\begin{tabular}{cccc}
\toprule
$l$ & $[[N,k]]$ & $N-k$ & $\chi_E$ \\
\midrule
2 & $[[4,1]]$ & 3 & 2 \\
3 & $[[8,4]]$ & 4 & 8 \\
4 & $[[16,11]]$ & 5 & 16 \\
5 & $[[32,26]]$ & 6 & 32 \\
6 & $[[64,57]]$ & 7 & 64 \\
7 & $[[128,120]]$ & 8 & 128 \\
\bottomrule
\end{tabular}
\end{center}

For $l \geq 3$ the encoding bond dimension saturates at $\chi_E = 2^l = N$.
The weight-$N$ logical $\bar{X}$ operator ($X^{\otimes N}$) is a useful heuristic indicator of why such linear-in-$N$ scaling is possible, because it guarantees support across every cut.
However, the exact statement $\chi_E = 2^l$ follows from evaluating the stabilizer-rank formula Eq.~\eqref{eq:chi_from_rank} on the full block matrix Eq.~\eqref{eq:G_from_code} established above, not from that row alone.
The small deviation at $l=2$ ($\chi_E = 2 < N = 4$) occurs because the code has only $k=1$ logical qubit, insufficient to saturate the entanglement.

This direct route enables efficient cost estimation: one can evaluate $\chi_E$ for any candidate CSS code from its check matrices alone, before committing to the circuit synthesis.

For the $l{=}5$ instance used throughout Sec.~\ref{sec:numerical_evidence}, the same calculation gives $\chi_E = 2^l = N = 32$ for the reference encoded plus state $\ket{\bar{+}}^{\otimes k}$. The corresponding logical circuit acts on $k=26$ qubits, so the logical bound of Eq.~\eqref{eq:logical_chi_bound} gives the balanced-cut upper bound $\chi_{\logi,\mathrm{max}} = 2^{13} = 8192$.
This is the exponential barrier that a direct MPS simulation of the logical circuit eventually runs into. If the dynamics scrambles, the logical bond dimension can grow toward $8192$, while the physical simulation stays pinned at $\chi_E=32$ for this reference input. The contrast between the linear-in-$N$ physical bound and the exponential-in-$k$ logical bound is the key quantitative gain from compiling through the code.

\section{Monomial logical circuits and alternative physical backend}
\label{sec-monomial}

The previous section simulated the compiled physical circuit as an MPS, at a cost set by the encoder bond dimension $\chi_E$. This is one of several possible backends, each with its own cost bottleneck. Here we describe a complementary exact algebraic backend. For a restricted class of circuits, the evolution can be tracked symbolically, at a cost controlled not by any bond dimension but by the degree of the phase polynomial.

The package \textsc{PhasePoly.jl}~\cite{phasepoly} implements the exact algebraic backend. It evaluates a restricted monomial subfamily by storing a binary bit-string map, a shift, and an accumulated phase polynomial. The goal of this section is to explain when this finite phase-sum backend is cheap, when it loses a general polynomial-time guarantee, and why the compiled physical circuit can still be useful for higher-level Clifford-hierarchy entries, including level $t=4$ entries.

The relevant subfamily is the \emph{monomial} circuits. A unitary is monomial in a basis if it maps each basis vector to one basis vector multiplied by a phase, or equivalently if its matrix has exactly one nonzero entry in every row and every column~\cite{vandennest2011monomial}. In such a circuit, a computational-basis string does not branch into a superposition at intermediate times. It follows a reversible classical trajectory and accumulates a phase. This is similar to the representation used in phase-polynomial treatments of optimization frameworks, for example in $T$-depth or $\CNOT$-depth optimizations~\cite{amy2014tdepth, chen2025phasepolyoptimizationframeworkforphase}. Here it is used as an exact simulator for some monomial circuits generated by the CCQC catalog.

For a single block, the monomial form is
\begin{equation}\label{eq-monomial-block}
B_r\ket{x} = e^{i p_r(x)}\ket{A_r x\oplus b_r},
\end{equation}
where $x=(x_1,\ldots,x_k)$ is a bit string, all additions are modulo $2$, $A_r$ is an invertible binary matrix, $b_r$ is a binary shift, and $p_r(x)$ is the phase accumulated by that block. A circuit made from such blocks remains monomial. Composing the blocks gives one final binary map with shift $(A,b)$ and one accumulated phase polynomial $p(x)$.

The diagonal catalog entries are the simplest monomial blocks. They have $A_r=I$ and $b_r=0$, so they leave the bit string fixed and only add a phase. This phase may be Clifford, such as $Z$, $S$, or $\CZ$, or non-Clifford, such as $T$, $\CS$, $\CCZ$, and, for validated level-$4$ entries, phases such as $\CT$, $\CCS$, or $\CCCZ$. All of these diagonal gates are monomial because they act as
\[
\ket{x}\longmapsto e^{ip_r(x)}\ket{x} .
\]

Automorphism entries can also fit this form. The direct phase-polynomial evaluator keeps the automorphisms whose induced action preserves the computational basis. Examples include maps such as $X$, $\CNOT$, and $\SWAP$, possibly accompanied by diagonal Clifford phases such as $Z$, $S$, and $\CZ$. These are the basis-preserving Clifford actions described by linear and quadratic functions over binary variables~\cite{dehaene2003}. Automorphisms with Hadamard-type action are different. They send a computational-basis state to a superposition, so one binary map plus one phase polynomial is no longer enough. This is a limitation of the algebraic backend, not of the CCQC construction. The compiled MPS simulator still handles it with the MPS cost bound above.

For the dense reference input $\ket{+}^{\otimes k}=2^{-k/2}\sum_x\ket{x}$, the resulting state is
\begin{equation}\label{eq-ppstate}
\ket{\psi}=2^{-k/2}\sum_{x} e^{ip(x)}\ket{Ax\oplus b} .
\end{equation}
Because $A$ is invertible, each output bit string comes from a unique input bit string. We can therefore relabel the sum by the output string and absorb this relabeling into a new phase function $q$,
\begin{equation}\label{eq-ppstate-output}
\ket{\psi}
=2^{-k/2}\sum_{y\in\{0,1\}^k} e^{iq(y)}\ket{y} .
\end{equation}
Here $q(y)$ is the phase attached to the unique input bit string that produces $y$.
Every amplitude has modulus $2^{-k/2}$, so computational-basis sampling is uniform and every nontrivial $Z$-only expectation value vanishes. The phase information is visible in $X$-, $Y$-, and mixed-Pauli observables. With the convention
$X_aZ_c\ket{y}=(-1)^{c\cdot y}\ket{y\oplus a}$, and with $Y_j=iX_jZ_j$, Eq.~\eqref{eq-ppstate-output} gives the finite-difference sum
\begin{equation}\label{eq-pauli-finite-diff}
\begin{aligned}
\langle X_aZ_c\rangle
&= \frac{1}{2^{k}}
\sum_{y\in\{0,1\}^k}
(-1)^{c\cdot y}\,
e^{\,i\,[q(y)-q(y\oplus a)]}.
\end{aligned}
\end{equation}
Thus direct expectation values are controlled by finite differences of the phase polynomial, not only by the polynomial itself. For the level-$3$ circuits considered here, taking the finite difference lowers the relevant phase polynomial from cubic to quadratic. The resulting sum can therefore be evaluated efficiently using the Gauss-sum method implemented in \textsc{PhasePoly.jl}~\cite{bu2022halfgauss,phasepoly}. It is the reduction to a quadratic (degree-two) phase that makes the evaluation efficient; the level-$4$ phases discussed below leave a cubic difference and fall outside this regime.

In the purely diagonal case, $U_f\ket{y}=(-1)^{f(y)}\ket{y}$ with a Boolean polynomial $f(y)\in\{0,1\}$, this reduces to
\begin{equation}
\begin{aligned}
\langle X_j\rangle
&=2^{-k}\sum_{y\in\{0,1\}^k} (-1)^{f(y)+f(y\oplus e_j)}\\
&=2^{-k}\operatorname{gap}(\Delta_j f),
\end{aligned}
\label{eq-x-gap}
\end{equation}
where $\Delta_j f(y)=f(y)-f(y\oplus e_j)$ modulo $2$. Since subtraction and addition are the same over binary variables, the exponent can equivalently be written as $f(y)+f(y\oplus e_j)$. The gap is a simple signed count,
\begin{equation}
\begin{aligned}
\operatorname{gap}(g)
&=\sum_y(-1)^{g(y)}\\
&=\#\{y\mid g(y)=0\}-\#\{y\mid g(y)=1\}.
\end{aligned}
\end{equation}
The circuit-polynomial correspondence from  ~\cite{montanaro2017lowdegree} uses this same quantity for $H,Z,\CZ,\CCZ$ circuits, and states that computing $\operatorname{gap}(f)$ for degree-$3$ polynomials is \#P-hard in general.

These finite differences also show where the direct evaluator loses its general efficiency. Some level-$4$ phases still leave low-degree differences, but genuinely degree-$4$ phase terms do not. For example, a logical $\CCCZ$ term contributes a Boolean phase $x_jx_ax_bx_c$. Taking the $X_j$ finite difference leaves $x_ax_bx_c$, and Eq.~\eqref{eq-x-gap} becomes the gap of a cubic Boolean polynomial. Therefore, although $\CCCZ$ does not break the phase-polynomial description, it breaks the general efficient quadratic-evaluation route for direct logical $X$-type expectations.

Special instances can still be easy to simulate. For example, logical circuits containing only level-3 gates $(t=3)$ and acting on stabilizer input states can be simulated efficiently using \textsc{PhasePoly.jl} \cite{phasepoly}. This efficiency can be lost when higher-level non-stabilizer resources introduce more complicated phase polynomials. 
Similarly, a sparse state-vector simulators stores only the nonzero computational-basis amplitudes and can be efficient when that support remains small~\cite{veroni2025}. The dense plus input used for the logical phase-sum calculation is the opposite regime. If $m$ independent leading plus variables are present, the support already has size $2^m$. When $m$ is large, support-based simulation is not the cheap description.

CCQC demonstrates that the compiled physical circuit can be a different and sometimes better phase-polynomial representation of the same logical operation. After encoding, every diagonal catalog entry, whether level $3$ or level $4$, is a product of single-qubit physical phases. Every automorphism entry is represented physically by a qubit relabeling while inducing a binary logical map. For a stabilizer-type input preparation, the physical phase-polynomial form is
\begin{equation}\label{eq-phys-form}
V_{\phys}=P_{\mathrm{diag}}\,A_{\mathrm{bin}}\,H_{\mathrm{lead}},
\end{equation}
where $H_{\mathrm{lead}}$ prepares $m$ leading plus variables, $A_{\mathrm{bin}}$ is the collapsed $\CNOT$/permutation/translation map, and $P_{\mathrm{diag}}$ is a product of single-qubit phases such as $Z$, $S$, $T$, and, for level-$4$ physical entries, $P(\pi/8)=\sqrt T$. 

If the compiled physical circuit contains only supported monomial gates, including single-qubit $\sqrt T$ phases, \textsc{PhasePoly.jl} can evaluate it exactly. With physical phases up to level $3$ it uses the quadratic Gauss-sum route. Once single-qubit $\sqrt T$ phases are present, the implementation switches to branch enumeration. The exponential cost then comes from the number of prepared branches.

One can therefore choose the backend for the compiled physical circuit based on the required resources. \textsc{PhasePoly.jl} is useful when the compiled circuit remains in the monomial, leading-Hadamard form and the branch count or residual phase structure is manageable. This includes product stabilizer or phase inputs and some entangled inputs prepared by monomial gates. For more general encoded initial states, or for circuits with non-monomial blocks, this algebraic representation is no longer the right backend. The tensor-network backend remains the main simulator whenever the encoded MPS bond dimension is manageable, because onsite physical phases and tracked relabelings do not increase that bond dimension.

In summary, the two backends are controlled by different complexity parameters. \textsc{PhasePoly.jl} is preferable when the input admits a leading-H, monomial preparation and the subsequent circuit consists of supported permutations, $\CNOT$s, and diagonal phases. For physical phases at level $\leq3$, each Pauli expectation reduces to a quadratic Gauss sum evaluable in polynomial time, independently of the state's entanglement. This advantage is weakened by level-$4$ phases: a Pauli expectation involving $t$ surviving $\sqrt T$ parity terms requires exponential evaluations. The MPS backend is therefore preferable whenever the encoded bond dimension remains moderate but the input is a generic product or MPS state, the circuit contains non-monomial onsite rotations, the physical phase layers are dense or extend beyond the supported hierarchy, or many observables and samples must be extracted from the same evolved state. Conversely, when the MPS bond dimension becomes large while the phase polynomial remains quadratic, \textsc{PhasePoly.jl} is preferable. In the monomial level-$3$ regime both backends apply and provide an independent cross-check.

\section{Quantum resource measures}
\label{sec:magic}

To quantify the resource content generated by these circuits, we track two complementary measures: nonstabilizerness (magic) and non-Gaussianity through the interaction distance from the entanglement spectrum. Together they capture different ways in which the evolving state departs from the standard efficiently simulable regimes reviewed above. Related work also connects nonstabilizerness to entanglement-spectrum flatness, giving a complementary way to probe magic in spectral structure beyond entropy alone~\cite{tirrito_flatness_2024}.

\subsection{Entanglement entropy and bond dimension}

At every circuit step $\tau$ we track two quantities. The first is the maximal cut entropy $S_{\max}(\tau):=\max_{A|B} S(\rho_A(\tau))$, where $S(\rho_A) = -\sum_i \lambda_i \log_2 \lambda_i$ is the von Neumann entropy of the reduced state across the bipartition $A|B$. The second is the largest bond dimension actually stored during the MPS simulation, $\chi_{\max}(\tau)$, which determines the runtime cost. The compilation mechanism shows up most directly in this pair: $\chi_{\max}$ stays pinned at $\chi_E$ for the compiled physical circuit, while it can grow substantially for the direct logical simulation.

\subsection{Interaction distance (non-Gaussianity)}
\label{sec:interaction}
Matchgate (free-fermion) simulation, reviewed in Sec.~\ref{sec:matchgate}, applies to states whose reduced density matrices have a Gaussian spectrum. To measure how far a state departs from this regime, we use the interaction distance~\cite{turner2017, pachosSciPost, matos_emergence_2021, deger_persistent_2023},
\begin{equation}
D_F(\rho_A) = \min_{\sigma} \frac{1}{2}\|\rho_A - \sigma\|_1,
\end{equation}
where the minimisation is over the set of density matrices with free-fermion (Gaussian) entanglement spectra, i.e.\ spectra generated by free single-particle entanglement energies. Therefore, $D_F$ depends only on the eigenvalues of $\rho_A$. A state with $D_F = 0$ has the same reduced spectrum as some Gaussian state, while a nonzero value signals genuine many-body interactions beyond the free-fermion description. Since $D_F$ is a trace distance between density matrices, it is bounded. For fermionic spectra the maximal value has been conjectured to be $D_F^{\max}=3-2\sqrt{2}$~\cite{deger_persistent_2023}. Values approaching this scale are therefore close to maximally non-Gaussian in the interaction-distance sense. This diagnostic is complementary to matchgate resource-theoretic results, where pure non-Gaussian fermionic states are precisely the resource states that promote matchgate computation beyond its Gaussian sector~\cite{hebenstreit_strelchuk_2019}.

For an MPS of bond dimension $\chi$, the reduced density matrix $\rho_A$ across a bipartition is obtained by a singular value decomposition at the cut, and its eigenvalues are the squares of the singular values. We evaluate $D_F$ at every bipartition position $i = 1, \ldots, N-1$ and at every circuit step $\tau$, producing a profile $D_F(\tau,i)$ that we visualize as a heatmap to show where and when non-Gaussianity develops.

\subsection{Nonstabilizerness estimation via perfect Pauli sampling}

The nonstabilizerness (magic) of a pure state $\ket{\psi}$ on $N$ qubits can be quantified by the stabilizer R\'{e}nyi entropies~\cite{leone2022}. Define the Pauli measurement distribution
\begin{equation}
\Pi_\sigma := \frac{1}{2^N}|\!\bra{\psi}\sigma\ket{\psi}\!|^2, \quad \sigma \in \mathcal{P}_N,
\end{equation}
which is a probability distribution over the $4^N$ Pauli strings. The order-$\alpha$ stabilizer R\'{e}nyi entropy is
\begin{equation}
M_\alpha := \frac{1}{1-\alpha}\log_2\!\left(\sum_{\sigma\in\mathcal{P}_N}\Pi_\sigma^\alpha\right) - N,
\end{equation}
which vanishes for stabilizer states and grows with nonstabilizerness. We use the normalised (per-qubit) density $m_\alpha = M_\alpha/N$.

To estimate $m_\alpha$ from an MPS representation of $\ket{\psi}$ we use the perfect Pauli sampling algorithm of Lami and Collura~\cite{lami2023}, which draws Pauli strings $\sigma$ exactly from $\Pi_\sigma$ via a sequential left-to-right sweep through the MPS. Given $S$ independent samples $\{(\sigma^{(\mu)},\Pi^{(\mu)})\}_{\mu=1}^S$, the order-1 density and its standard error are
\begin{equation}
m_1 = -\frac{\langle\log_2 \Pi\rangle}{N} - 1, \qquad \delta m_1 = \frac{\mathrm{std}(\log_2 \Pi)}{\sqrt{S}\,N},
\end{equation}
where $\langle\cdot\rangle$ denotes the sample mean.

\section{Numerical results}
\label{sec:numerical_evidence}
We test the CCQC simulation framework using the polar instance at $l{=}5$, namely the $[[32,26]]$ code. We generate random circuits from the matched gate library, with expanded logical circuit gate count $L\sim 300$, apply them to simple product inputs ($\ket{+}^{\otimes k}$ and random product states), and simulate them with MPS using a truncation cutoff $\epsilon = 10^{-10}$. For each circuit we run two simulations in parallel: the logical circuit directly on $k$ qubits, and the compiled physical circuit on $N$ qubits with classical permutation tracking. When comparing the two simulation paths we report agreement at the numerical tolerance set by this truncation cutoff, and use exact/no-truncation checks on small instances to verify the relabeling implementation. The Figures below summarize the main diagnostics.

\begin{figure}[t!]
\centering
\includegraphics[width=0.95\linewidth]{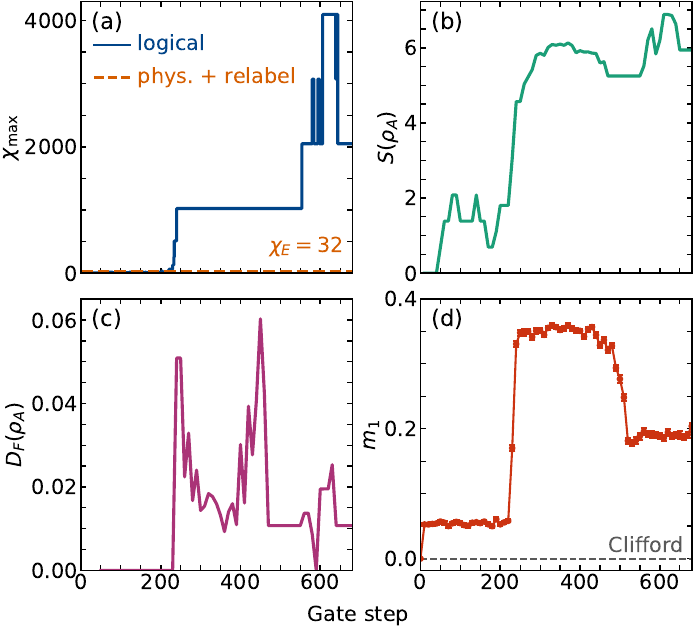}
\caption{Representative resource diagnostics for the polar $l=5$ $[[32,26]]$ code family. \textbf{(a)}~The physical simulation with classical relabeling stays pinned at the encoding cost $\chi_E=32$, while the direct logical simulation develops large bond-dimension spikes. \textbf{(b)}~The maximal cut entropy grows substantially with depth. \textbf{(c)}~The interaction distance shows departure from the free-fermion manifold. \textbf{(d)}~The nonstabilizerness density $m_1$ rises well above the Clifford baseline.}
\label{fig:resources}
\end{figure}

The central numerical result is shown in Fig.~\ref{fig:resources}. The figure reports the resource diagnostics generated by representative random circuits from the matched catalog. The circuit logical gates are dense, highly nonlocal operations yet each compiles to a short physical circuit consisting only of onsite phases and a few permutations. Figure~\ref{fig:resources}(a) then shows the dynamical consequence: the physical simulation with classical relabeling stays pinned to $\chi_E$, while the direct logical simulation develops large, growing bond-dimension spikes that would eventually exhaust any fixed computational budget. Figure~\ref{fig:resources}(b) shows that this is not a low-entanglement regime: the maximal cut entropy grows substantially.
Figure~\ref{fig:resources}(c) shows that the reduced states also leave the free-fermion manifold, since the interaction distance $D_F$~\cite{pachosSciPost,matos_emergence_2021,deger_persistent_2023} reaches appreciable values across multiple cuts.

The benchmark expectation values and correlation functions computed from the physical MPS agree with those of the logical simulation to the $10^{-10}$ truncation tolerance, confirming that the on-the-fly relabeling reproduces the logical circuit correctly. Figure~\ref{fig:resources}(d) shows the same pattern for nonstabilizerness. It rises well above stabilizer baselines under perfect Pauli sampling~\cite{leone2022,lami2023}.
Taken together, these diagnostics show that the logical circuit looks increasingly hard from standard entanglement, magic, and Gaussianity proxies, while the compiled physical simulation remains controlled by the fixed encoding bond dimension. Across the polar family, the same qualitative behavior persists as $l$ increases: the physical bond dimension after encoding scales polynomially with $N$ while the logical simulation develops large depth-dependent spikes.

\begin{figure}[t!]
\centering
\includegraphics[width=0.95\linewidth]{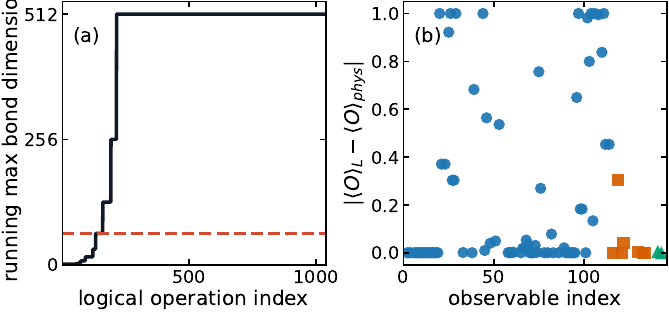}
\caption{Simulation for the polar $l=6$ $[[64,57]]$ code family. \textbf{(a)}~The solid black curve is the direct logical MPS running over 656 logical operations with maximum allowed bond dimension $\tilde{\chi}=512$. The dashed red curve is the compiled physical simulation, which uses a 64-qubit MPS. The physical bond dimension remains fixed at 64. \textbf{(b)}~For each observable $O$, we compare the capped logical MPS to the compiled physical reference using $\Delta_O=|\langle \bar{O}\rangle-\langle O\rangle_{\phys}|$. Blue circles mark all single-qubit $X_i$ and $Z_i$ observables on the 57 logical qubits, orange squares mark a sampled set of ten $XX$ and ten $ZZ$ pairs, and green triangles mark ten sampled four-$X$ correlators.}
\label{fig:l6_logical_cap}
\end{figure}

The larger $l=6$ instance in Fig.~\ref{fig:l6_logical_cap} demonstrates the practical consequence of this separation. We simulate a circuit on $k=57$ logical qubits encoded into $N=64$ physical qubits. In the compiled physical representation, all post-encoding operations are onsite phases and relabelings, so the MPS bond dimension remains fixed at the encoder value $\chi_E=64=N$. By contrast, the direct logical simulation reaches the imposed cap $\tilde{\chi}=512$ partway through the circuit, as shown in Fig.~\ref{fig:l6_logical_cap}(a). Beyond that point, the capped logical state no longer reproduces the compiled-reference observables. The discrepancies in Fig.~\ref{fig:l6_logical_cap}(b) should therefore be interpreted as truncation error from the direct logical simulation, not as a failure of the compiled physical evolution. This is precisely the regime in which the physical representation is useful: it continues to provide a controlled classical reference after the direct tensor-network simulation has exhausted its fixed bond-dimension budget.

\section{Application: Hardware benchmarking via logical direct fidelity estimation}
\label{sec:dfe_application}

Near-term and early fault-tolerant quantum devices will need to be benchmarked at the level of logical circuit performance, rather than relying solely on the calibration of individual physical gates. CCQCs
provide a natural route to such a benchmark: the hardware runs a long-range,
non-Clifford circuit \(U\), while the classically simulated compiled
MPS is used to predict selected observables of its ideal logical output.

Our compiled reference provides the ideal Pauli expectation values required by direct fidelity estimation (DFE)~\cite{flammia2011}, making standard DFE applicable to the non-Clifford, long-range logical circuits considered here. Let
\[
\ket{\psi}=U \ket{\psi_{\rm in}},
\]
be the ideal \(k\)-qubit output state. Let \(\sigma\) denote the corresponding
\(k\)-qubit state produced by the device when it runs \(U\) on the same input $\ket{\psi_{\rm in}}$. For each Pauli
string \(Q\in\mathcal{P}_k\), define
\begin{equation}
 r_Q=\bra{\psi}Q\ket{\psi},\qquad s_Q=\Tr(\sigma Q),\qquad
 w(Q)=2^{-k}r_Q^{\,2}.
\label{eq:w_L_def}
\end{equation}
Purity of \(\ket{\psi}\) implies \(\sum_Q r_Q^{\,2}=2^k\), so \(w\) is a probability distribution. Expanding \(\ket{\psi}\!\bra{\psi}\) and \(\sigma\) in the Pauli basis, and using \(\Tr(QQ')=2^k\delta_{Q,Q'}\), gives
\begin{equation}
 F=\bra{\psi}\sigma\ket{\psi}
 =\mathbb{E}_{Q\sim w}\!\left[\frac{s_Q}{r_Q}\right],
\label{eq:logical_dfe}
\end{equation}
which is the logical fidelity between the ideal and device output states.

Operationally, DFE draws \(Q\) from the target-dependent distribution
\(w\), estimates \(s_Q\) on repeated device outputs, and averages
\(s_Q/r_Q\).  The device therefore measures selected logical Pauli
observables rather than reconstructing a full output distribution: it
contributes \(s_Q\), while the compiled MPS selects \(Q\) and predicts \(r_Q\).

\begin{figure}[t!]
  \includegraphics[width=0.95\linewidth]{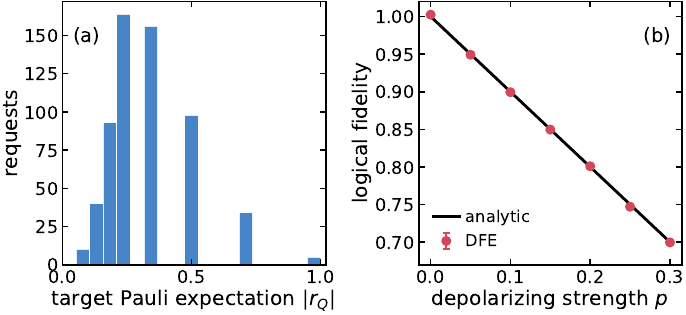}
  \caption{Logical-DFE worked example for the polar $l=4$,
  $[[16,11]]$ CCQC instance. \textbf{(a)}~Distribution of the magnitudes
  of the target logical Pauli expectations
  $|r_Q|=|\bra{\psi}Q\ket{\psi}|$ over the $600$ sampled non-identity measurement
  requests at the 32-layer circuit.  \textbf{(b)}~Estimator validation
  on a simulated noisy device.  A depolarizing channel
  $\sigma=(1-p)\ket{\psi}\!\bra{\psi}+pI/2^{k}$ is applied to the logical output for
  $p\in[0,0.30]$. The black line is the analytic state fidelity
  $F(p)=(1-p)+p/2^{k}$ and the red markers are DFE estimates from
  simulated finite-shot measurements of the requested $Q$ strings
  ($30$ estimates per $p$, $400$ shots per request.)}
  \label{fig:dfe_l4}
\end{figure}

The CCQC is suitable for two reasons. First, the MPS backend gives efficient access to the ideal Pauli expectation values required for direct fidelity estimation. Second, the coherent phases introduced by the non-Clifford layers may be invisible in simple population measurements but are exposed by \(X\)-, \(Y\)-, and mixed-Pauli correlations. The benchmark therefore probes whether the device has reproduced the logical interference structure of \(U\), rather than merely whether individual gates or computational-basis populations look correct.

As a quick illustration, if the target is the Bell state \(\ket{\Phi^+}\), then the only nonzero coefficients are
\[
r_{II}=r_{XX}=r_{ZZ}=1,\qquad r_{YY}=-1,
\]
each with relevance weight \(1/4\). Given measured values
\[
s_{XX}=0.92,\qquad s_{ZZ}=0.95,\qquad s_{YY}=-0.88,
\]
the estimator returns
\begin{align}
\widehat F
&=
\frac14\left[
\frac{s_{II}}{r_{II}}
+
\frac{s_{XX}}{r_{XX}}
+
\frac{s_{ZZ}}{r_{ZZ}}
+
\frac{s_{YY}}{r_{YY}}
\right] \notag
\\
&=
\frac14\left[
1
+
\frac{0.92}{1}
+
\frac{0.95}{1}
+
\frac{-0.88}{-1}
\right]
=
0.9375,
\end{align}
with the identity term contributing the known value \(s_{II}/r_{II}=1\).

The non-trivial step is obtaining samples from \(w\) and computing the signed
ideal values \(r_Q\) without directly simulating a large target circuit. Rather
than sampling \(w\) directly, the classical backend samples from the compiled
physical representation. Let \(a=N-k\) and
\[
\ket{\Psi_E}=E\bigl(\ket{0}^{\otimes a}\otimes\ket{\psi}\bigr).
\]
Perfect Pauli sampling on the compiled MPS draws \(P\in\mathcal P_N\) with
\[
W(P)=2^{-N}\bigl\langle \Psi_E \big| P \big| \Psi_E\bigr\rangle^2.
\]
Because \(E\) is Clifford, the sampled physical Pauli can be pulled back as
\[
E^\dagger P E=\eta(P)A(P)\otimes Q(P),
\]
where \(A(P)\in\mathcal P_a\) acts on the auxiliary qubits and
\(Q(P)\in\mathcal P_k\) is the Pauli requested from the device.
The auxiliary expectation vanishes unless \(A(P)\) contains only \(I\)'s and
\(Z\)'s. For each fixed logical \(Q\), the nonzero auxiliary branches have
total weight \(2^a2^{-N}r_Q^2=2^{-k}r_Q^2=w(Q)\). Thus sampling \(P\) from the
compiled MPS and keeping \(Q(P)\) samples the correct logical distribution.
For each retained sample, the signed ideal value is obtained from the same MPS expectation, since
\[
r_{Q(P)}=\eta(P)\ \langle\Psi_E|P|\Psi_E\rangle
\]
on the nonzero auxiliary branches.

The clearest way to see this is to write down the smallest non-trivial
case explicitly.  Take $k=1$, $a=1$, $E=\CNOT_{L, A}$, order the
tensor factors as \(A\otimes L\), and use the target
\(\ket{\psi}=\ket{+}\).  Then \(\ket{\Psi_E}=\ket{\Phi^+}\), and the four Pauli strings with
nonzero $W$, together with their pushbacks, are

\begin{center}
\setlength{\tabcolsep}{8pt}
\begin{tabular}{@{}cccccc@{}}
\toprule
$P$ & $E^\dagger P E$ & $\eta$ & $A$ & $Q$ & $W(P)$\\
\midrule
$I \otimes I$ & $\phantom{+}I\otimes I$  & $+$ & $I$ & $I$ & $1/4$\\
$Z \otimes Z$ & $\phantom{+}Z\otimes I$  & $+$ & $Z$ & $I$ & $1/4$\\
$X \otimes X$ & $\phantom{+}I\otimes X$  & $+$ & $I$ & $X$ & $1/4$\\
$Y \otimes Y$ & $-Z\otimes X$            & $-$ & $Z$ & $X$ & $1/4$\\
\bottomrule
\end{tabular}
\end{center}
Here the two rows returning \(Q=I\) have total probability \(1/2\), and the
two rows returning \(Q=X\) also have total probability \(1/2\).  This is
exactly the DFE distribution of the logical target \(\ket{+}\).
Thus the compiled MPS chooses the logical measurement \(Q(P)\) and supplies its
ideal value \(r_{Q(P)}\), whereas the hardware only measures that logical Pauli
after running \(U\).

We now demonstrate this benchmark on the polar $[[16,11]]$ CCQC
instance, for which $N=16$, $k=11$, and $a=5$.  The logical circuit is
a sequence drawn from the $l=4$ catalog.
Each catalog layer is one automorphism (Clifford) or transversal-diagonal (non-Clifford) block of the logical circuit, and we examine circuits containing $32$ such layers.
At $l=4$ the encoded reference can be constructed exactly, which allows the
ingredients of the protocol to be cross-checked against a direct
logical evaluation. The compiled-physical and encoded-logical outputs
agree in state fidelity to numerical precision.

Let's say a representative output of the classical reference path for this
circuit is
\[
 Q=Z_1X_2I_3I_4X_5Y_6X_7Z_8I_9Z_{10}I_{11},
 \qquad r_Q=0.25.
\]
The corresponding hardware operation is a readout of $\sigma$ in the
local Pauli basis specified by $Q$: logical qubits $1$, $8$, $10$ are
measured in the $Z$ basis, qubits $2$, $5$, $7$ in the $X$ basis,
qubit $6$ in the $Y$ basis, and the identity positions are not measured.
The product of the $\pm1$ outcomes over each shot is one realisation
of $Q$, and the sample mean over many shots is $\widehat s_Q$.  An
observed value $\widehat s_Q=0.20$, for instance, contributes
$\widehat s_Q/r_Q=0.80$ to the DFE average.

Figure~\ref{fig:dfe_l4}(a) shows the distribution of $|r_Q|$ over the
$600$ non-identity measurement requests sampled at the 32-layer
circuit. Sampling replaces the formal
$4^k\approx 4.2\times 10^{6}$-term Pauli expansion of $F$ by an
unbiased average over sampled requests. The required number of requests depends on the desired accuracy, finite-shot noise, and the variance induced by the sampled values. In this worked example we use $600$ requests as a representative finite-sample demonstration to estimate the fidelity of this $11$-qubit logical output.
The presence of weight strictly below $|r_Q|=1$ reflects the
non-Clifford structure of the target state: a stabilizer-state target
would have all relevance weight at $|r_Q|=1$, while a Haar-random
target on $k=11$ qubits would concentrate near
$|r_Q|\sim 2^{-k/2}\approx 0.022$.  The polar instance lies between
these extremes, consistent with the diagonal non-Clifford layers
generating a structured but nontrivial Pauli spectrum.

Figure~\ref{fig:dfe_l4}(b) validates the estimator end-to-end against
a controlled noise model.  We replace the device output by a
depolarized logical state $\sigma=(1-p)\ket{\psi}\!\bra{\psi}+pI/2^k$ for
$p\in[0,0.30]$ and compute its analytic state fidelity,
\begin{equation}
 F(p)=(1-p)+p/2^{k},
\label{eq:depol_fid}
\end{equation}
shown as the black curve.  The red markers are DFE estimates obtained
from simulated finite-shot measurements of the requested $Q$ strings,
with $30$ independent estimates per $p$ and $400$ shots per
non-identity request. The identity contribution is included
analytically.  The estimates track $F(p)$ within their statistical
error bars across the full range.

In this construction the compiled MPS provides both the measurement
schedule $\{Q_i\}$ and the ideal answers $\{r_{Q_i}\}$, while the
hardware provides the measured answers $\{\widehat s_{Q_i}\}$. Their
ratio average
$\widehat F=\frac{1}{M}\sum_{i=1}^{M}\widehat s_{Q_i}/r_{Q_i}$ is a
global state-fidelity estimate for the output of $U$. Selected
fixed logical correlators can be measured in parallel.

\section{Discussion}
\label{sec:discussion}

We have introduced code-compiled quantum circuits, a class of quantum circuits constructed by compiling a logical circuit through a CSS encoder so that every layer after encoding consists of single-qubit diagonal gates and a permutation of qubit labels. The logical dynamics of the construction contains both Clifford and non-Clifford gates, accepts a broad input class including states with magic or non-Gaussianity, and can be made arbitrarily deep. The accompanying classical simulation cost is governed entirely by the peak encoder bond dimension $\chi_E$, with all subsequent layers contributing only single-qubit and classical-relabeling operations. For the polar CSS family and the encoded reference stabilizer state, this bond dimension grows linearly with the number of physical qubits, while the logical bond dimension would grow exponentially in the number of encoded qubits, and the gap widens with circuit depth. Large entanglement, nonstabilizerness, and non-Gaussianity are therefore not sufficient indicators of classical hardness on their own.

Beyond simulation, our framework enables a hardware-benchmarking protocol that applies to NISQ and early fault-tolerant devices. A target circuit that is deep, long-range, and non-Clifford typically admits no efficient classical reference, and the error a device incurs while running it cannot be estimated directly. CCQCs offer a matched benchmark of comparable profile --- similar depth, similar density of non-Clifford gates, similar logical connectivity --- for which an efficient classical reference exists by construction. Running the CCQC on the device and comparing against this reference yields a sample-efficient estimate of how well the hardware handles operations of that style, and the estimate can be used as a proxy for the fidelity of the original target.

We have demonstrated the protocol using direct fidelity estimation~\cite{flammia2011} on the polar instance. Perfect Pauli sampling on the compiled MPS, followed by a Clifford pushback through the encoder, yields both the measurement schedule and the ideal expectation values, so that the device runs the logical circuit and performs only local Pauli readout. This provides an end-to-end benchmark for deep, non-Clifford circuits while retaining an efficiently computable ideal reference. As a result, the benchmark scales naturally to depths and qubit counts, up to the usual sample complexity of DFE~\cite{flammia2011}. The same idea applies to any code admitting an efficient encoder and a rich library of automorphism and transversal diagonal gates.

Structurally, the CCQC encoder plays the role of a physical transformation that relates a strongly correlated logical evolution to a physical evolution whose only nontrivial action is onsite, in the similar spirit as the Jordan--Wigner transformation between interacting spin systems and free fermions or holographic codes between bulk and boundary degrees of freedom~\cite{almheiri2015,pastawski2015}. The encoded state stores the entanglement that the logical circuit would have to build dynamically. Our construction therefore draws an explicit connection between quantum error correction~\cite{calderbank1996,steane1996} and classical simulability~\cite{aaronson2004,schollwock2011,jozsa2008}, adding CCQCs to the families of efficiently simulable quantum circuits.

Several directions remain open. The polar family studied here is one instance of a wider construction, and identifying CSS codes with richer automorphism groups, more usable transversal diagonal gates, or favorable $\chi_E$ scaling is a natural next step.
The MPS and \textsc{PhasePoly.jl}~\cite{phasepoly} backends are not the only viable classical approaches once the compiled structure is exposed: extended-Clifford and stabilizer-decomposition methods~\cite{bravyi2016,bravyi2018cliffordt}, Pauli-propagation approaches~\cite{rall2019,aharonov2023,angrisani2024}, and approximate tensor-network contractions such as belief propagation~\cite{tindall2023} are also candidates. On the logical side, the code-determined gate library is inherently non-universal by the Eastin--Knill theorem~\cite{eastin2009}, and characterising the expressibility and computational power of the gate sets that the construction does produce is an open question.

{\bf \em Acknowledgments:--}
AD acknowledges support from the EPSRC through the QCi3 Hub (EP/Z53318X/1). AD and DEB acknowledge support from the QCS Hub (EP/T001062/1), which supported the early stages of this work. JR is funded by an EPSRC Quantum Career Acceleration Fellowship (grant code: UKRI1224). JR and SK were supported by the Innovate UK project \enquote{QEC Readout Testbed} [reference number 10151107] and an EPSRC IAA Cross Institutional Project between University of Edinburgh and University of Glasgow. HS is supported by the EPSRC [grant number EP/S021582/1]. HS also acknowledges support from the National Physical Laboratory. MW is supported by the EPSRC [grant number EP/W032635/1 and EP/S005021/1] and  Innovate UK [grant number 10179725].
\bibliographystyle{apsrev4-2}
\bibliography{references}

@article{Dias2024classicalsimulation,
  doi = {10.22331/q-2024-05-21-1350},
  url = {https://doi.org/10.22331/q-2024-05-21-1350},
  title = {Classical simulation of non-{G}aussian fermionic circuits},
  author = {Dias, Beatriz and Koenig, Robert},
  journal = {{Quantum}},
  issn = {2521-327X},
  publisher = {{Verein zur F{\"{o}}rderung des Open Access Publizierens in den Quantenwissenschaften}},
  volume = {8},
  pages = {1350},
  month = may,
  year = {2024}
}

@misc{chen2025phasepolyoptimizationframeworkforphase,
      title={Leveraging Phase Polynomials for Quantum Circuit Optimization}, 
      author={Zihan Chen and Henry Chen and Yuwei Jin and Enhyeok Jang and Mingkuan Xu and Vannessa Chan and Won Woo Ro and Eddy Z. Zhang},
      year={2026},
      eprint={2506.20624},
      archivePrefix={arXiv},
      primaryClass={cs.PL},
      url={https://arxiv.org/abs/2506.20624}, 
}

@article{bravyi2018cliffordt,
  doi = {10.22331/q-2019-09-02-181},
  url = {https://doi.org/10.22331/q-2019-09-02-181},
  title = {Simulation of quantum circuits by low-rank stabilizer decompositions},
  author = {Bravyi, Sergey and Browne, Dan and Calpin, Padraic and Campbell, Earl and Gosset, David and Howard, Mark},
  journal = {{Quantum}},
  issn = {2521-327X},
  publisher = {{Verein zur F{\"{o}}rderung des Open Access Publizierens in den Quantenwissenschaften}},
  volume = {3},
  pages = {181},
  month = sep,
  year = {2019}
}

@article{nahum2017,
	title = {Quantum Entanglement Growth under Random Unitary Dynamics},
	author = {Nahum, Adam and Ruhman, Jonathan and Vijay, Sagar and Haah, Jeongwan},
	journal = {Physical Review X},
	volume = {7},
	pages = {031016},
	year = {2017},
	doi = {10.1103/PhysRevX.7.031016},
}

@article{angrisani2024,
  title         = {Classically estimating observables of noiseless quantum circuits},
  author        = {Angrisani, Armando and Schmidhuber, Alexander and Rudolph, Manuel S. and Cerezo, M. and Holmes, Zo{\"e} and Huang, Hsin-Yuan},
  journal       = {Physical Review Letters},
  volume        = {135},
  pages         = {170602},
  year          = {2025},
  doi           = {10.1103/lh6x-7rc3},
  eprint        = {2409.01706},
  archivePrefix = {arXiv},
  primaryClass  = {quant-ph},
}

@article{flammia2011,
  title   = {Direct Fidelity Estimation from Few Pauli Measurements},
  author  = {Flammia, Steven T. and Liu, Yi-Kai},
  journal = {Physical Review Letters},
  volume  = {106},
  pages   = {230501},
  year    = {2011},
  doi     = {10.1103/PhysRevLett.106.230501},
}

@article{sayginel2024,
	title = {Fault-tolerant logical {C}lifford gates from code automorphisms},
	author = {Sayginel, Hasan and Koutsioumpas, Stergios and Webster, Mark and Rajput, Abhishek and Browne, Dan E.},
	journal = {PRX Quantum},
	volume = {6},
	pages = {030343},
	year = {2025},
	doi = {10.1103/vf7v-cpq9},
	url = {https://doi.org/10.1103/vf7v-cpq9},
}

@article{calderbank1996,
	title = {Good quantum error-correcting codes exist},
	author = {Calderbank, A. R. and Shor, Peter W.},
	journal = {Physical Review A},
	volume = {54},
	pages = {1098--1105},
	year = {1996},
	doi = {10.1103/PhysRevA.54.1098},
}

@article{steane1996,
	title = {Error correcting codes in quantum theory},
	author = {Steane, A. M.},
	journal = {Physical Review Letters},
	volume = {77},
	pages = {793--797},
	year = {1996},
	doi = {10.1103/PhysRevLett.77.793},
}

@misc{fattal2004,
	title = {Entanglement in the Stabilizer Formalism},
	author = {Fattal, David and Cubitt, Toby S. and Yamamoto, Yoshihisa and Bravyi, Sergey and Chuang, Isaac L.},
	year = {2004},
	doi = {10.48550/arXiv.quant-ph/0406168},
	eprint = {quant-ph/0406168},
	archivePrefix = {arXiv},
	primaryClass = {quant-ph},
	url = {https://arxiv.org/abs/quant-ph/0406168},
}

@article{veroni2025,
	title = {Universal Quantum Computation via Scalable Measurement-Free Error Correction},
	author = {Veroni, Stefano and Paler, Alexandru and Giudice, Giacomo},
	journal = {PRX Quantum},
	volume = {6},
	number = {4},
	pages = {040337},
	year = {2025},
	doi = {10.1103/lkk1-v6wp},
	url = {https://journals.aps.org/prxquantum/abstract/10.1103/lkk1-v6wp},
	eprint = {2412.15187},
	archivePrefix = {arXiv},
	primaryClass = {quant-ph},
}

@article{bombin2006,
	title = {Topological quantum distillation},
	author = {Bomb{\'\i}n, H. and Martin-Delgado, M. A.},
	journal = {Physical Review Letters},
	volume = {97},
	pages = {180501},
	year = {2006},
	doi = {10.1103/PhysRevLett.97.180501},
}

@article{vasmer2019,
	title = {Three-dimensional surface codes: Transversal gates and fault-tolerant architectures},
	author = {Vasmer, Michael and Browne, Dan E.},
	journal = {Physical Review A},
	volume = {100},
	pages = {012312},
	year = {2019},
	doi = {10.1103/PhysRevA.100.012312},
	url = {https://journals.aps.org/pra/abstract/10.1103/PhysRevA.100.012312},
}

@article{menendez2024,
	title = {Implementing fault-tolerant non-Clifford gates using the [[8,3,2]] color code},
	author = {Honciuc Menendez, Daniel and Ray, Annie and Vasmer, Michael},
	journal = {Physical Review A},
	volume = {109},
	pages = {062438},
	year = {2024},
	doi = {10.1103/PhysRevA.109.062438},
	url = {https://journals.aps.org/pra/abstract/10.1103/PhysRevA.109.062438},
}

@misc{mocherla2023,
	title = {Extending matchgate simulation methods to universal quantum circuits},
	author = {Mocherla, Avinash and Lao, Lingling and Browne, Dan E.},
	year = {2023},
	doi = {10.48550/arXiv.2302.02654},
	eprint = {2302.02654},
	archivePrefix = {arXiv},
	primaryClass = {quant-ph},
	url = {https://arxiv.org/abs/2302.02654},
}

@inproceedings{gottesman1998,
	author = {Gottesman, Daniel},
	title = {The Heisenberg Representation of Quantum Computers},
	booktitle = {Group22: Proceedings of the XXII International Colloquium on Group Theoretical Methods in Physics},
	pages = {32--43},
	year = {1999},
	publisher = {International Press},
	doi = {10.48550/arXiv.quant-ph/9807006},
	eprint = {quant-ph/9807006},
	archivePrefix = {arXiv},
	url = {https://arxiv.org/abs/quant-ph/9807006},
	note = {Also available as arXiv:quant-ph/9807006},
}

@article{dehaene2003,
	title = {Clifford Group, Stabilizer States, and Linear and Quadratic Operations over GF(2)},
	author = {Dehaene, Jeroen and De Moor, Bart},
	journal = {Physical Review A},
	volume = {68},
	pages = {042318},
	year = {2003},
	doi = {10.1103/PhysRevA.68.042318},
}

@article{aaronson2004,
	title = {Improved Simulation of Stabilizer Circuits},
	author = {Aaronson, Scott and Gottesman, Daniel},
	journal = {Physical Review A},
	volume = {70},
	pages = {052328},
	year = {2004},
	doi = {10.1103/PhysRevA.70.052328},
	eprint = {quant-ph/0406196},
	archivePrefix = {arXiv},
}

@article{valiant2002,
	title = {Quantum circuits that can be simulated classically in polynomial time},
	author = {Valiant, Leslie G.},
	journal = {SIAM Journal on Computing},
	volume = {31},
	number = {4},
	pages = {1229--1254},
	year = {2002},
	doi = {10.1137/S0097539700377025},
	url = {https://doi.org/10.1137/S0097539700377025},
}

@article{terhal2002,
	title = {Classical simulation of noninteracting-fermion quantum circuits},
	author = {Terhal, Barbara M. and DiVincenzo, David P.},
	journal = {Physical Review A},
	volume = {65},
	pages = {032325},
	year = {2002},
	doi = {10.1103/PhysRevA.65.032325},
}

@article{jozsa2008,
	title = {Matchgates and classical simulation of quantum circuits},
	author = {Jozsa, Richard and Miyake, Akimasa},
	journal = {Proceedings of the Royal Society A},
	volume = {464},
	number = {2100},
	pages = {3089--3106},
	year = {2008},
	doi = {10.1098/rspa.2008.0189},
	url = {https://doi.org/10.1098/rspa.2008.0189},
}

@article{vidal2003,
	title = {Efficient Classical Simulation of Slightly Entangled Quantum Computations},
	author = {Vidal, Guifr{\'e}},
	journal = {Physical Review Letters},
	volume = {91},
	pages = {147902},
	year = {2003},
	doi = {10.1103/PhysRevLett.91.147902},
	eprint = {quant-ph/0301063},
	archivePrefix = {arXiv},
}

@article{schollwock2011,
	title = {The density-matrix renormalization group in the age of matrix product states},
	author = {Schollw{\"o}ck, Ulrich},
	journal = {Annals of Physics},
	volume = {326},
	number = {1},
	pages = {96--192},
	year = {2011},
	doi = {10.1016/j.aop.2010.09.012},
	eprint = {1008.3477},
	archivePrefix = {arXiv},
}

@article{eastin2009,
	title = {Restrictions on Transversal Encoded Quantum Gate Sets},
	author = {Eastin, Bryan and Knill, Emanuel},
	journal = {Physical Review Letters},
	volume = {102},
	pages = {110502},
	year = {2009},
	doi = {10.1103/PhysRevLett.102.110502},
	eprint = {0811.4262},
	archivePrefix = {arXiv},
}

@article{lami2023,
	title = {Nonstabilizerness via perfect Pauli sampling of matrix product states},
	author = {Lami, Guglielmo and Collura, Mario},
	journal = {Physical Review Letters},
	volume = {131},
	pages = {180401},
	year = {2023},
	doi = {10.1103/PhysRevLett.131.180401},
}

@article{turner2017,
	title = {Optimal free descriptions of many-body theories},
	author = {Turner, C. J. and Meichanetzidis, K. and Papi{\'c}, Z. and Pachos, J. K.},
	journal = {Nature Communications},
	volume = {8},
	pages = {14926},
	year = {2017},
	doi = {10.1038/ncomms14926},
}

@Article{pachosSciPost,
	title={{Quantifying the effect of interactions in quantum many-body systems}},
	author={Jiannis K. Pachos and Zlatko Papic},
	journal={SciPost Phys. Lect. Notes},
	pages={4},
	year={2018},
	publisher={SciPost},
	doi={10.21468/SciPostPhysLectNotes.4},
	url={https://scipost.org/10.21468/SciPostPhysLectNotes.4},
}

@article{bravyi2005,
	title = {Universal quantum computation with ideal Clifford gates and noisy ancillas},
	author = {Bravyi, Sergey and Kitaev, Alexei},
	journal = {Physical Review A},
	volume = {71},
	pages = {022316},
	year = {2005},
	doi = {10.1103/PhysRevA.71.022316},
}

@article{gottesman_chuang1999,
	title = {Demonstrating the Viability of Universal Quantum Computation Using Teleportation and Single-Qubit Operations},
	author = {Gottesman, Daniel and Chuang, Isaac L.},
	journal = {Nature},
	volume = {402},
	pages = {390--393},
	year = {1999},
	doi = {10.1038/46503},
}

@article{bravyi2016,
	title = {Improved classical simulation of quantum circuits dominated by Clifford gates},
	author = {Bravyi, Sergey and Gosset, David},
	journal = {Physical Review Letters},
	volume = {116},
	pages = {250501},
	year = {2016},
	doi = {10.1103/PhysRevLett.116.250501},
}

@inproceedings{junttila2007,
	title = {Engineering an efficient canonical labeling tool for large and sparse graphs},
	author = {Junttila, Tommi and Kaski, Petteri},
	booktitle = {Proceedings of the Ninth Workshop on Algorithm Engineering and Experiments (ALENEX 2007)},
	pages = {135--149},
	year = {2007},
	doi = {10.1137/1.9781611972870.13},
	url = {https://doi.org/10.1137/1.9781611972870.13},
}

@article{leone2022,
	title = {Stabilizer R{\'e}nyi entropy},
	author = {Leone, Lorenzo and Oliviero, Salvatore F. E. and Hamma, Alioscia},
	journal = {Physical Review Letters},
	volume = {128},
	pages = {050402},
	year = {2022},
	doi = {10.1103/PhysRevLett.128.050402},
}

@article{rall2019,
	title = {Simulation of qubit quantum circuits via Pauli propagation},
	author = {Rall, Patrick and Liang, Daniel and Cook, Jeremy and Kretschmer, William},
	journal = {Physical Review A},
	volume = {99},
	pages = {062337},
	year = {2019},
	doi = {10.1103/PhysRevA.99.062337},
}

@inproceedings{aharonov2023,
	title = {A polynomial-time classical algorithm for noisy random circuit sampling},
	author = {Aharonov, Dorit and Gao, Xun and Landau, Zeph and Liu, Yunchao and Vazirani, Umesh},
	booktitle = {Proceedings of the 55th Annual ACM Symposium on Theory of Computing},
	pages = {945--957},
	year = {2023},
	doi = {10.1145/3564246.3585234},
	url = {https://doi.org/10.1145/3564246.3585234},
}

@article{hayden2007,
	title = {Black holes as mirrors: quantum information in random subsystems},
	author = {Hayden, Patrick and Preskill, John},
	journal = {Journal of High Energy Physics},
	volume = {2007},
	number = {09},
	pages = {120},
	year = {2007},
	doi = {10.1088/1126-6708/2007/09/120},
}

@article{sekino2008,
	title = {Fast scramblers},
	author = {Sekino, Yasuhiro and Susskind, Leonard},
	journal = {Journal of High Energy Physics},
	volume = {2008},
	number = {10},
	pages = {065},
	year = {2008},
	doi = {10.1088/1126-6708/2008/10/065},
}

@article{almheiri2015,
	title = {Bulk locality and quantum error correction in {AdS}/{CFT}},
	author = {Almheiri, Ahmed and Dong, Xi and Harlow, Daniel},
	journal = {Journal of High Energy Physics},
	volume = {2015},
	number = {04},
	pages = {163},
	year = {2015},
	doi = {10.1007/JHEP04(2015)163},
}

@article{pastawski2015,
	title = {Holographic quantum error-correcting codes: toy models for the bulk/boundary correspondence},
	author = {Pastawski, Fernando and Yoshida, Beni and Harlow, Daniel and Preskill, John},
	journal = {Journal of High Energy Physics},
	volume = {2015},
	number = {06},
	pages = {149},
	year = {2015},
	doi = {10.1007/JHEP06(2015)149},
}

@article{Webster_2023,
doi = {10.1088/1367-2630/acfc5f},
url = {https://iopscience.iop.org/article/10.1088/1367-2630/acfc5f/meta},
year = {2023},
month = {oct},
publisher = {IOP Publishing},
volume = {25},
number = {10},
pages = {103018},
author = {Webster, Mark A and Quintavalle, Armanda O and Bartlett, Stephen D},
title = {Transversal diagonal logical operators for stabiliser codes},
journal = {New Journal of Physics}
}

@article{cui2017,
	title = {Diagonal Gates in the Clifford Hierarchy},
	author = {Cui, Shawn X. and Gottesman, Daniel and Krishna, Anirudh},
	journal = {Physical Review A},
	volume = {95},
	pages = {012329},
	year = {2017},
	doi = {10.1103/PhysRevA.95.012329},
}

@article{bombin2015,
	title = {Gauge Color Codes: Optimal Transversal Gates and Gauge Fixing in Topological Stabilizer Codes},
	author = {Bomb{\'\i}n, H{\'e}ctor},
	journal = {New Journal of Physics},
	volume = {17},
	number = {8},
	pages = {083002},
	year = {2015},
	doi = {10.1088/1367-2630/17/8/083002},
}

@misc{hangleiter2026,
  title     = {Has quantum advantage been achieved?},
  author    = {Hangleiter, Dominik},
  year      = {2026},
  doi       = {10.48550/arXiv.2603.09901},
  eprint    = {2603.09901},
  archivePrefix = {arXiv},
  primaryClass  = {quant-ph},
  url       = {https://arxiv.org/abs/2603.09901},
}

@article{arikan2009channel,
  title     = {Channel Polarization: A Method for Constructing Capacity-Achieving Codes for Symmetric Binary-Input Memoryless Channels},
  author    = {Ar{\i}kan, Erdal},
  journal   = {IEEE Transactions on Information Theory},
  volume    = {55},
  number    = {7},
  pages     = {3051--3073},
  year      = {2009},
  doi       = {10.1109/TIT.2009.2021379},
  eprint    = {0807.3917},
  archivePrefix = {arXiv},
}

@article{renes2012efficient,
  title     = {Efficient Polar Coding of Quantum Information},
  author    = {Renes, Joseph M. and Dupuis, Fr{\'e}d{\'e}ric and Renner, Renato},
  journal   = {Physical Review Letters},
  volume    = {109},
  number    = {5},
  pages     = {050504},
  year      = {2012},
  doi       = {10.1103/PhysRevLett.109.050504},
  eprint    = {1109.3195},
  archivePrefix = {arXiv},
}

@article{wilde2013polar,
  title     = {Polar Codes for Classical-Quantum Channels},
  author    = {Wilde, Mark M. and Guha, Saikat},
  journal   = {IEEE Transactions on Information Theory},
  volume    = {59},
  number    = {2},
  pages     = {1175--1187},
  year      = {2013},
  doi       = {10.1109/TIT.2012.2218792},
  eprint    = {1109.2591},
  archivePrefix = {arXiv},
}

@misc{gottesman_stabilizer_1997,
    title = {Stabilizer {Codes} and {Quantum} {Error} {Correction}},
    url = {http://arxiv.org/abs/quant-ph/9705052},
    doi = {10.48550/arXiv.quant-ph/9705052},
    urldate = {2024-06-29},
    publisher = {arXiv},
    author = {Gottesman, Daniel},
    month = may,
    year = {1997},
    note = {arXiv:quant-ph/9705052},
}

@misc{gottesman_surviving_nodate,
    title = {Surviving as a {Quantum} {Computer} in a {Classical} {World}, 2024 {Draft}},
    url = {https://www.cs.umd.edu/class/spring2024/cmsc858G/QECCbook-2024-ch1-15.pdf},
    author = {Gottesman, Daniel},
    year = {2024},
}

@inproceedings{grassl_leveraging_2013,
    title = {Leveraging automorphisms of quantum codes for fault-tolerant quantum computation},
    url = {https://ieeexplore.ieee.org/document/6620283},
    doi = {10.1109/ISIT.2013.6620283},
    urldate = {2024-06-07},
    booktitle = {2013 {IEEE} {International} {Symposium} on {Information} {Theory}},
    author = {Grassl, Markus and Roetteler, Martin},
    month = jul,
    year = {2013},
    pages = {534--538},
}

@misc{hao_investigations_2021,
    title = {Investigations on {Automorphism} {Groups} of {Quantum} {Stabilizer} {Codes}},
    url = {http://arxiv.org/abs/2109.12735},
    doi = {10.48550/arXiv.2109.12735},
    urldate = {2024-06-07},
    publisher = {arXiv},
    author = {Hao, Hanson},
    month = sep,
    year = {2021},
    note = {arXiv:2109.12735 [cs, math]},
}

@article{jain_high-distance_2024,
     author={Jain, Shubham P. and Albert, Victor V.},
  journal={IEEE Journal on Selected Areas in Information Theory}, 
  title={Transversal Clifford and T-Gate Codes of Short Length and High Distance}, 
  year={2025},
  volume={6},
  pages={127--137},
  doi={10.1109/JSAIT.2025.3570832}}

@misc{koutsioumpas_smallest_2022,
    title = {The {Smallest} {Code} with {Transversal} {T}},
    url = {http://arxiv.org/abs/2210.14066},
    doi = {10.48550/arXiv.2210.14066},
    urldate = {2024-11-19},
    publisher = {arXiv},
    author = {Koutsioumpas, Stergios and Banfield, Darren and Kay, Alastair},
    month = oct,
    year = {2022},
    note = {arXiv:2210.14066},
}

@article{PhysRevLett.120.050504,
  title = {Distillation with Sublogarithmic Overhead},
  author = {Hastings, Matthew B. and Haah, Jeongwan},
  journal = {Phys. Rev. Lett.},
  volume = {120},
  issue = {5},
  pages = {050504},
  numpages = {3},
  year = {2018},
  month = {Jan},
  publisher = {American Physical Society},
  doi = {10.1103/PhysRevLett.120.050504},
  url = {https://link.aps.org/doi/10.1103/PhysRevLett.120.050504}
}

@article{PhysRevA.86.052329,
  title = {Magic-state distillation with low overhead},
  author = {Bravyi, Sergey and Haah, Jeongwan},
  journal = {Phys. Rev. A},
  volume = {86},
  issue = {5},
  pages = {052329},
  numpages = {10},
  year = {2012},
  month = {Nov},
  publisher = {American Physical Society},
  doi = {10.1103/PhysRevA.86.052329},
  url = {https://link.aps.org/doi/10.1103/PhysRevA.86.052329}
}

@article{PhysRevA.54.4741,
  title = {Simple quantum error-correcting codes},
  author = {Steane, A. M.},
  journal = {Phys. Rev. A},
  volume = {54},
  issue = {6},
  pages = {4741--4751},
  numpages = {0},
  year = {1996},
  month = {Dec},
  publisher = {American Physical Society},
  doi = {10.1103/PhysRevA.54.4741},
  url = {https://link.aps.org/doi/10.1103/PhysRevA.54.4741}
}

@ARTICLE{681315,
  author={Calderbank, A.R. and Rains, E.M. and Shor, P.W. and Sloane, N.J.A.},
  journal={IEEE Transactions on Information Theory}, 
  title={Quantum error correction via codes over GF(4)}, 
  year={1998},
  volume={44},
  number={4},
  pages={1369--1387},
  doi={10.1109/18.681315}}

@ARTICLE{7208851,
  author={Renes, Joseph M. and Sutter, David and Dupuis, Fr{\'e}d{\'e}ric and Renner, Renato},
  journal={IEEE Transactions on Information Theory}, 
  title={Efficient Quantum Polar Codes Requiring No Preshared Entanglement}, 
  year={2015},
  volume={61},
  number={11},
  pages={6395--6414},
  doi={10.1109/TIT.2015.2468084}}

@article{PhysRevLett.113.030501,
  title = {Tensor Networks and Quantum Error Correction},
  author = {Ferris, Andrew J. and Poulin, David},
  journal = {Phys. Rev. Lett.},
  volume = {113},
  issue = {3},
  pages = {030501},
  numpages = {5},
  year = {2014},
  month = {Jul},
  publisher = {American Physical Society},
  doi = {10.1103/PhysRevLett.113.030501},
  url = {https://link.aps.org/doi/10.1103/PhysRevLett.113.030501}
}

@INPROCEEDINGS{10619465,
  author={Gong, Anqi and Renes, Joseph M.},
  booktitle={2024 IEEE International Symposium on Information Theory (ISIT)}, 
  title={Improved Logical Error Rate via List Decoding of Quantum Polar Codes}, 
  year={2024},
  volume={},
  number={},
  pages={2496--2501},
  doi={10.1109/ISIT57864.2024.10619465}}

@ARTICLE{TannerG,
  author={Tanner, R.},
  journal={IEEE Transactions on Information Theory}, 
  title={A recursive approach to low complexity codes}, 
  year={1981},
  volume={27},
  number={5},
  pages={533--547},
  doi={10.1109/TIT.1981.1056404}}

@article{MCKAY201494,
title = {Practical graph isomorphism, II},
journal = {Journal of Symbolic Computation},
volume = {60},
pages = {94--112},
year = {2014},
issn = {0747-7171},
doi = {10.1016/j.jsc.2013.09.003},
url = {https://www.sciencedirect.com/science/article/pii/S0747717113001193},
author = {McKay, Brendan D. and Piperno, Adolfo}
}

@article{Barenco_univ,
  title = {Elementary gates for quantum computation},
  author = {Barenco, Adriano and Bennett, Charles H. and Cleve, Richard and DiVincenzo, David P. and Margolus, Norman and Shor, Peter and Sleator, Tycho and Smolin, John A. and Weinfurter, Harald},
  journal = {Phys. Rev. A},
  volume = {52},
  issue = {5},
  pages = {3457--3467},
  numpages = {0},
  year = {1995},
  month = {Nov},
  publisher = {American Physical Society},
  doi = {10.1103/PhysRevA.52.3457},
  url = {https://link.aps.org/doi/10.1103/PhysRevA.52.3457}
}

@article{webster_xp_2022,
    title = {The {XP} {Stabiliser} {Formalism}: a {Generalisation} of the {Pauli} {Stabiliser} {Formalism} with {Arbitrary} {Phases}},
    volume = {6},
    shorttitle = {The {XP} {Stabiliser} {Formalism}},
    url = {https://quantum-journal.org/papers/q-2022-09-22-815/},
    doi = {10.22331/q-2022-09-22-815},
    language = {en-GB},
    urldate = {2024-08-06},
    journal = {Quantum},
    author = {Webster, Mark A. and Brown, Benjamin J. and Bartlett, Stephen D.},
    month = sep,
    year = {2022},
    pages = {815},
}

@article{matos_emergence_2021,
    title = {Emergence of gaussianity in the thermodynamic limit of interacting fermions},
    volume = {104},
    issn = {2469-9950, 2469-9969},
    url = {https://link.aps.org/doi/10.1103/PhysRevB.104.L180408},
    doi = {10.1103/PhysRevB.104.L180408},
    language = {en},
    number = {18},
    urldate = {2024-05-01},
    journal = {Physical Review B},
    author = {Matos, Gabriel and Hallam, Andrew and Deger, Aydin and Papić, Zlatko and Pachos, Jiannis K.},
    month = nov,
    year = {2021},
    pages = {L180408},
}

@article{deger_persistent_2023,
    title = {Persistent {Non}-{Gaussian} {Correlations} in {Out}-of-{Equilibrium} {Rydberg} {Atom} {Arrays}},
    volume = {4},
    url = {https://link.aps.org/doi/10.1103/PRXQuantum.4.040339},
    doi = {10.1103/PRXQuantum.4.040339},
    language = {en},
    number = {4},
    urldate = {2024-05-01},
    journal = {PRX Quantum},
    author = {Deger, Aydin and Daniel, Aiden and Papić, Zlatko and Pachos, Jiannis K.},
    month = dec,
    year = {2023},
    pages = {040339},
}

@article{tindall2023,
    title = {Gauging tensor networks with belief propagation},
    volume = {15},
    issn = {2542-4653},
    url = {https://scipost.org/SciPostPhys.15.6.222},
    doi = {10.21468/SciPostPhys.15.6.222},
    number = {6},
    journal = {SciPost Physics},
    author = {Tindall, Joseph and Fishman, Matthew T.},
    year = {2023},
    pages = {222},
}

@article{daley_practical_2022,
    title = {Practical quantum advantage in quantum simulation},
    volume = {607},
    issn = {0028-0836, 1476-4687},
    url = {https://www.nature.com/articles/s41586-022-04940-6},
    doi = {10.1038/s41586-022-04940-6},
    language = {en},
    number = {7920},
    urldate = {2024-06-16},
    journal = {Nature},
    author = {Daley, Andrew J. and Bloch, Immanuel and Kokail, Christian and Flannigan, Stuart and Pearson, Natalie and Troyer, Matthias and Zoller, Peter},
    month = jul,
    year = {2022},
    pages = {667--676},
}

@article{vandennest2011monomial,
    title = {A monomial matrix formalism to describe quantum many-body states},
    author = {Van den Nest, Maarten},
    journal = {New Journal of Physics},
    volume = {13},
    number = {12},
    pages = {123004},
    year = {2011},
    doi = {10.1088/1367-2630/13/12/123004},
}

@article{amy2014tdepth,
    title = {Polynomial-time {T}-depth optimization of {Clifford}+{T} circuits via matroid partitioning},
    author = {Amy, Matthew and Maslov, Dmitri and Mosca, Michele},
    journal = {IEEE Transactions on Computer-Aided Design of Integrated Circuits and Systems},
    volume = {33},
    number = {10},
    pages = {1476--1489},
    year = {2014},
    doi = {10.1109/TCAD.2014.2341953},
    eprint = {1303.2042},
    archivePrefix = {arXiv},
    primaryClass = {quant-ph},
}

@article{bu2022halfgauss,
    title = {Classical simulation of quantum circuits by half {G}auss sums},
    author = {Bu, Kaifeng and Koh, Dax Enshan},
    journal = {Communications in Mathematical Physics},
    volume = {390},
    pages = {471--500},
    year = {2022},
    doi = {10.1007/s00220-022-04320-1},
    eprint = {1812.00224},
    archivePrefix = {arXiv},
    primaryClass = {quant-ph},
}

@article{montanaro2017lowdegree,
    title = {Quantum circuits and low-degree polynomials over {$\mathbb{F}_2$}},
    author = {Montanaro, Ashley},
    journal = {Journal of Physics A: Mathematical and Theoretical},
    volume = {50},
    number = {8},
    pages = {084002},
    year = {2017},
    doi = {10.1088/1751-8121/aa565f},
    eprint = {1607.08473},
    archivePrefix = {arXiv},
    primaryClass = {quant-ph},
}

@misc{phasepoly,
    title = {{PhasePoly.jl}: Fast exact simulation of monomial quantum circuits via phase polynomials},
    note = {Julia package, version 0.1.0},
    author = {Aydin Deger},
    howpublished = {\url{https://github.com/aydindeger/PhasePoly.jl}},
    year = {2026},
}

@misc{maslov_fast_2024,
      title={Fast classical simulation of Harvard/QuEra IQP circuits}, 
      author={Dmitri Maslov and Sergey Bravyi and Felix Tripier and Andrii Maksymov and Joe Latone},
      year={2024},
      eprint={2402.03211},
      archivePrefix={arXiv},
      primaryClass={quant-ph},
      url={https://arxiv.org/abs/2402.03211},
}

@misc{jozsa_miyake_strelchuk_2015,
      title={Jordan-Wigner formalism for arbitrary 2-input 2-output matchgates and their classical simulation}, 
      author={Richard Jozsa and Akimasa Miyake and Sergii Strelchuk},
      year={2015},
      eprint={1311.3046},
      archivePrefix={arXiv},
      primaryClass={quant-ph},
      url={https://arxiv.org/abs/1311.3046}, 
}

@article{hebenstreit_strelchuk_2019,
  title = {All Pure Fermionic Non-Gaussian States Are Magic States for Matchgate Computations},
  author = {Hebenstreit, M. and Jozsa, R. and Kraus, B. and Strelchuk, S. and Yoganathan, M.},
  journal = {Phys. Rev. Lett.},
  volume = {123},
  issue = {8},
  pages = {080503},
  numpages = {5},
  year = {2019},
  month = {Aug},
  publisher = {American Physical Society},
  doi = {10.1103/PhysRevLett.123.080503},
  url = {https://link.aps.org/doi/10.1103/PhysRevLett.123.080503}
}

@article{wahl_strelchuk_2023,
  title = {Simulating Quantum Circuits Using Efficient Tensor Network Contraction Algorithms with Subexponential Upper Bound},
  author = {Wahl, Thorsten B. and Strelchuk, Sergii},
  journal = {Phys. Rev. Lett.},
  volume = {131},
  issue = {18},
  pages = {180601},
  numpages = {6},
  year = {2023},
  month = {Oct},
  publisher = {American Physical Society},
  doi = {10.1103/PhysRevLett.131.180601},
  url = {https://link.aps.org/doi/10.1103/PhysRevLett.131.180601}
}

@misc{wille_strelchuk_2025,
      title={Classical simulation of parity-preserving quantum circuits}, 
      author={Carolin Wille and Sergii Strelchuk},
      year={2025},
      eprint={2504.19317},
      archivePrefix={arXiv},
      primaryClass={quant-ph},
      url={https://arxiv.org/abs/2504.19317},
}

@misc{oh2026fermionic,
      title={Classical simulation of free-fermionic dynamics and quantum chemistry with magic input}, 
      author={Changhun Oh and Michał Oszmaniec and Oliver Reardon-Smith and Zoltán Zimborás},
      year={2026},
      eprint={2604.26813},
      archivePrefix={arXiv},
      primaryClass={quant-ph},
      url={https://arxiv.org/abs/2604.26813}, 
}

@article{tirrito_flatness_2024,
  title = {Quantifying nonstabilizerness through entanglement spectrum flatness},
  author = {Tirrito, Emanuele and Tarabunga, Poetri Sonya and Lami, Gugliemo and Chanda, Titas and Leone, Lorenzo and Oliviero, Salvatore F. E. and Dalmonte, Marcello and Collura, Mario and Hamma, Alioscia},
  journal = {Phys. Rev. A},
  volume = {109},
  issue = {4},
  pages = {L040401},
  numpages = {6},
  year = {2024},
  month = {Apr},
  publisher = {American Physical Society},
  doi = {10.1103/PhysRevA.109.L040401},
  url = {https://link.aps.org/doi/10.1103/PhysRevA.109.L040401}
}

@article{raussendorf2003,
  title = {Measurement-based quantum computation on cluster states},
  author = {Raussendorf, Robert and Browne, Daniel E. and Briegel, Hans J.},
  journal = {Phys. Rev. A},
  volume = {68},
  issue = {2},
  pages = {022312},
  numpages = {32},
  year = {2003},
  month = {Aug},
  publisher = {American Physical Society},
  doi = {10.1103/PhysRevA.68.022312},
  url = {https://link.aps.org/doi/10.1103/PhysRevA.68.022312}
}

\appendix

\onecolumngrid
\clearpage
\section{Physical/logical pairs at level $t=4$}
\label{app:t4_examples}

\begin{figure}[h!]
    \centering
    \includegraphics[width=\linewidth]{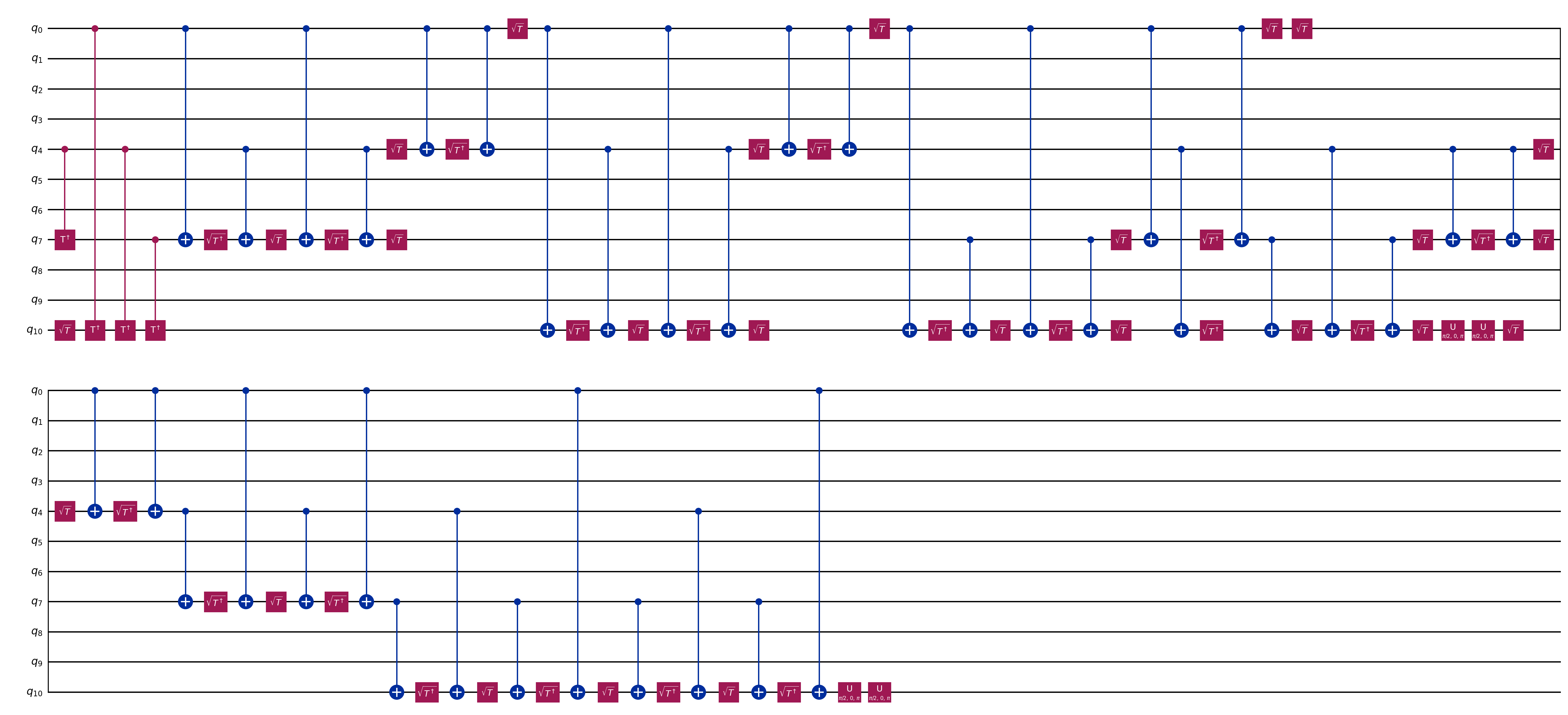}\\[2pt]
    \includegraphics[angle=-90,width=0.55\linewidth]{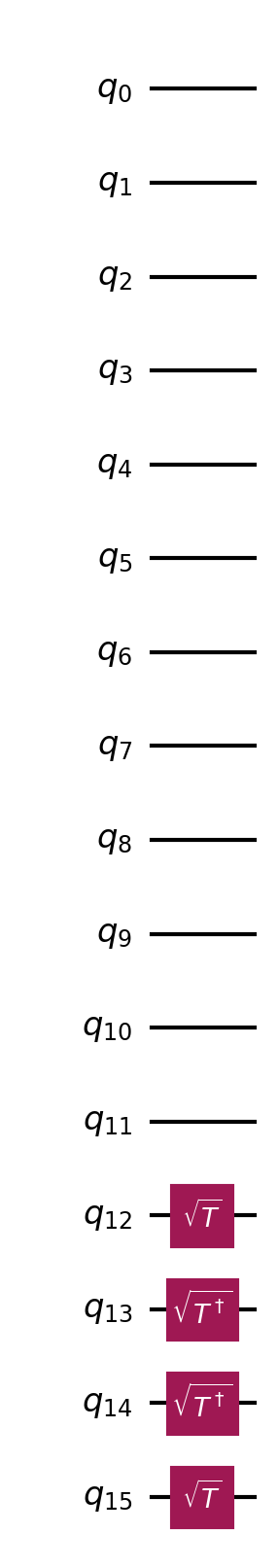}\\
    \caption{Matched pair from the level-$t=4$ catalog of the $[[16,11]]$ polar code.
    (a) Logical circuit: a $90$-gate non-Clifford circuit containing $\sqrt{T}$ and four $\mathrm{CT}^\dagger$ gates plus a Clifford+$\sqrt{T}$ synthesis.
    (b) Compiled physical circuit (drawn rotated by $90^\circ$ so the $N=16$ qubits run horizontally). No circuit optimisation has been applied.}
    \label{fig:t4_g011}
\end{figure}

\begin{figure}[h!]
    \centering
    \includegraphics[angle=-90,width=1\linewidth]{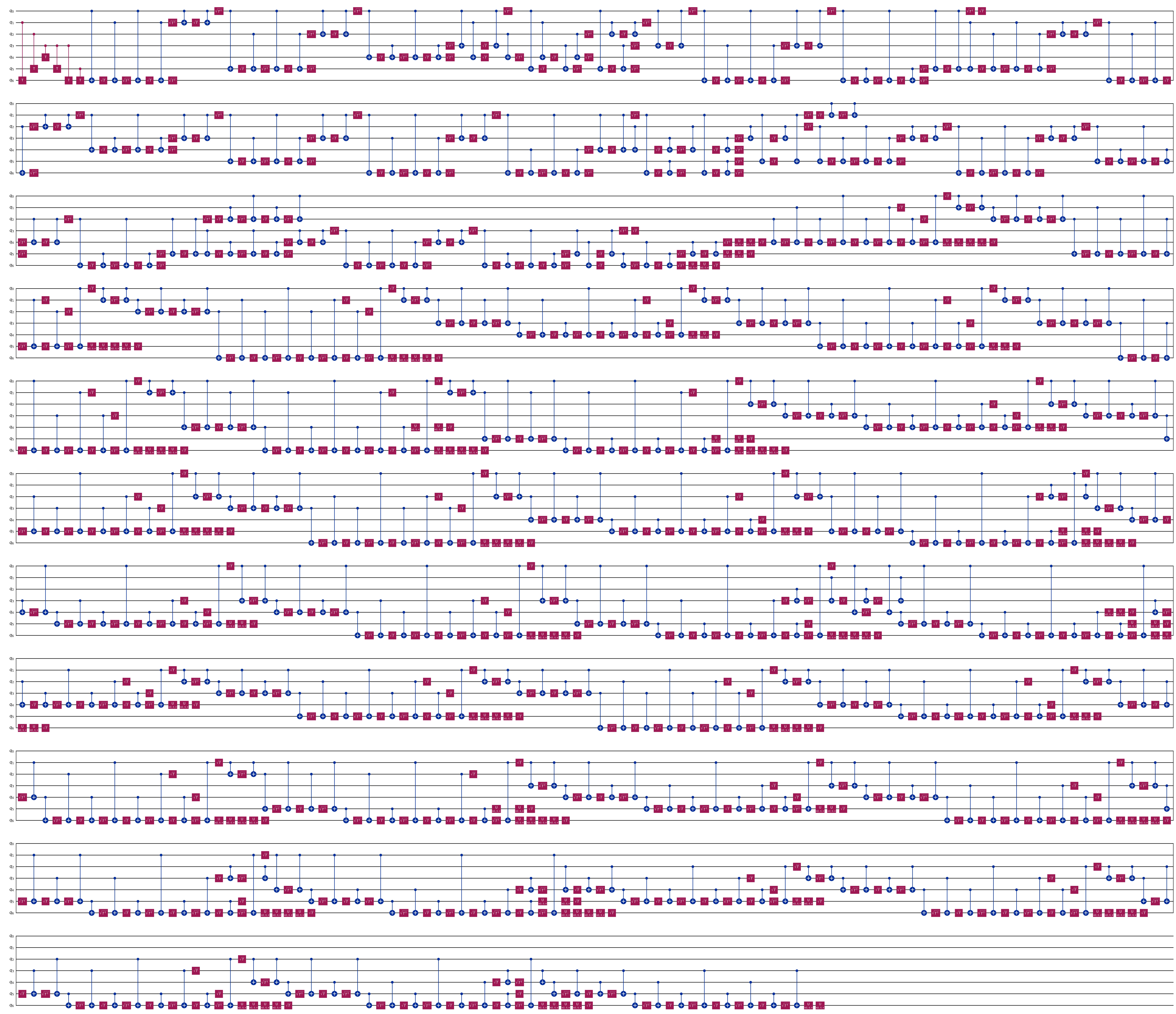}\\[2pt]
    \includegraphics[angle=-90,width=0.6\linewidth]{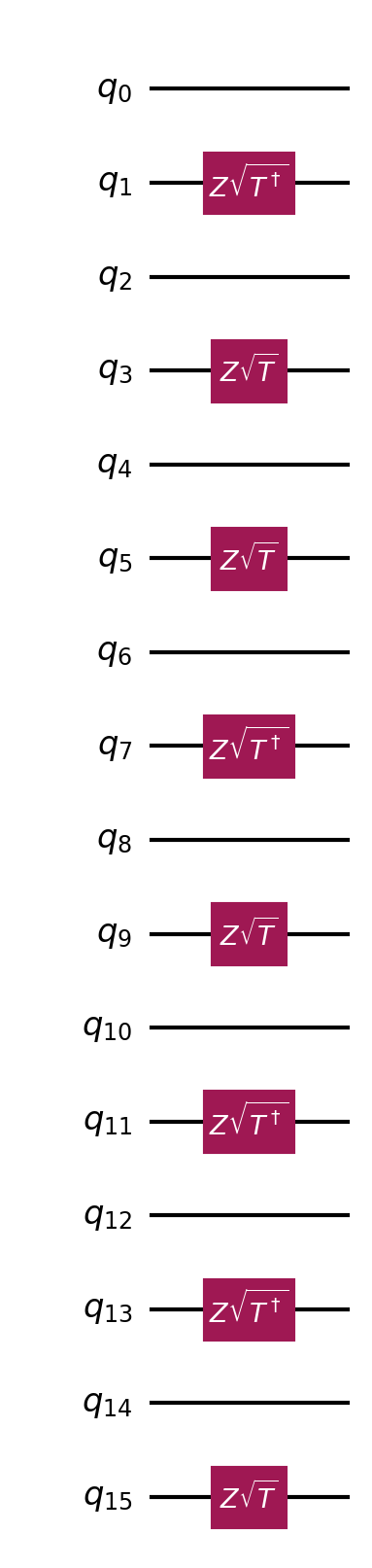}\\
    \caption{Matched pair from the level-$t=4$ catalog of the $[[16,11]]$ polar code.
    (a) Logical circuit: a $1361$-gate non-Clifford circuit.
    (b) Compiled physical circuit (drawn rotated by $90^\circ$ so the $N=16$ qubits run horizontally). No circuit optimisation has been applied.}
    \label{fig:t4_g012}
\end{figure}

We present here two representative matched pairs returned by the diagonal search at level $t=4$ for the $l{=}4$ polar instance, i.e.\ the $[[16,11]]$ code. At this level the physical alphabet contains the eighth-root-of-unity phase $P(\pi/8)=\sqrt{T}$ and its conjugate $\sqrt{T}^\dagger$, and the corresponding logical alphabet acquires gates from the fourth level of the Clifford hierarchy, including $\sqrt{T}$, the controlled-$T^\dagger$ gate (denoted $\mathrm{CT}^\dagger$), $\CCS$, and $\CCCZ$. We note that we decompose them using $\sqrt{T}$. Crucially, every compiled physical layer is still a product of single-qubit phases so the simulator cost does not grow when moving from level $3$ to level $4$. See Examples~A and~B (Figs.~\ref{fig:t4_g011} and~\ref{fig:t4_g012}) for the details of two such pairs.

\begin{figure}[h!]
    \centering
    \includegraphics[width=\linewidth]{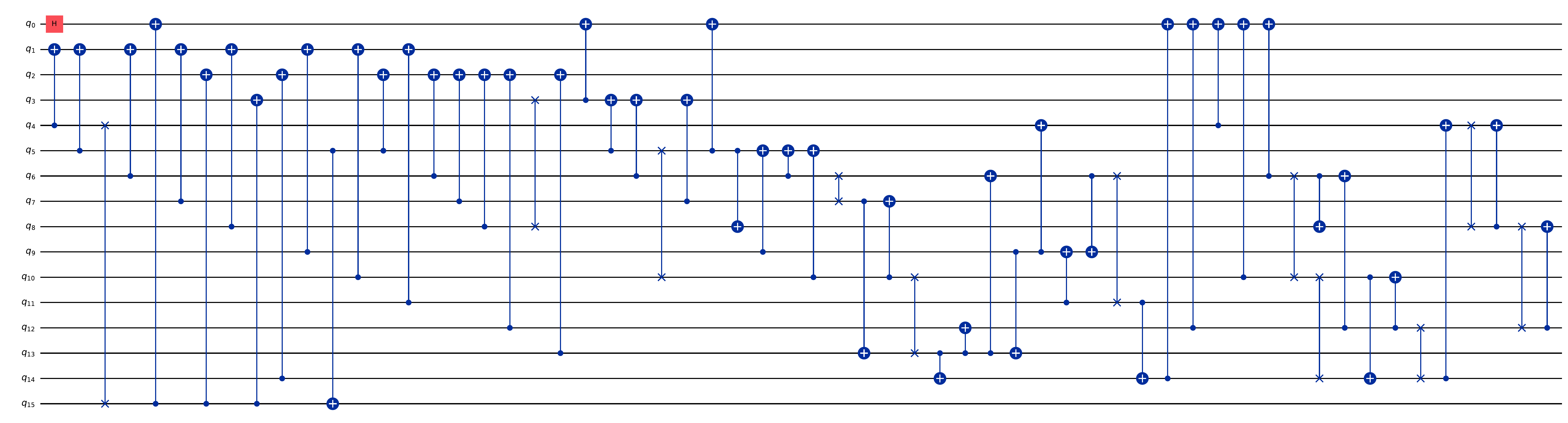}
    \caption{Encoder $E$ for the polar $[[16,11]]$ code ($l{=}4$), a Clifford circuit composed of $H$, $\CNOT$, and $\SWAP$ gates (qubits $q_0,\dots,q_{15}$). This is the only simulation stage that can grow the MPS bond dimension. Its peak value $\chi_E=N=16$ bounds the entire physical-side simulation cost. No circuit optimisation has been applied.}
    \label{fig:encoder_16_11}
\end{figure}

The encoder that compiles these logical circuits into their onsite physical form is shown in Fig.~\ref{fig:encoder_16_11}: a Clifford circuit ($H$, $\CNOT$, and $\SWAP$ gates). We verify the matched-pair equivalence of Fig.~\ref{fig:group}(d) directly on Example~A (the $G011$ pair of Fig.~\ref{fig:t4_g011}) by simulating it in both pictures. The logical circuit acts on $\ket{+}^{\otimes 11}$. In the \emph{physical} picture the $k=11$ logical qubits start in $\ket{+}$ and the $N-k=5$ auxiliary qubits in $\ket{0}$, the encoder $E$ is applied, and then the compiled physical layer. The physical bond dimension stays pinned at the encoder cost $\chi_E=N=16$, consistent with Eq.~\eqref{eq:cost_bound}.

\end{document}